\documentclass[a4paper]{article}
\usepackage{graphicx}
\usepackage{amsmath}
\usepackage{amssymb}
\usepackage[sort&compress,numbers]{natbib}
\usepackage{url}
\usepackage[a4paper, margin=1in]{geometry}
\usepackage{algorithm}
\usepackage{algpseudocode}
\begin{document}
% \linenumbers

\title{Indirect reciprocity under opinion synchronization}
\author{
  Yohsuke Murase\thanks{RIKEN Center for Computational Science, Kobe, 650-0047, Japan, \texttt{yohsuke.murase@gmail.com}} \and
  Christian Hilbe\thanks{Max Planck Research Group `Dynamics of Social Behavior', Max Planck Institute for Evolutionary Biology, Pl\"on, 24306, Germany}
}

\newcommand{\SI}{\textbf{SI}}
\newcommand{\meth}{\textbf{Methods}}

\newcommand{\YM}[1]{\textcolor{blue}{\textbf{YM}: #1}}
\newcommand{\christian}[1]{\textcolor{blue}{\textbf{CH}: #1}}
\newcommand{\lp}[1]{\left( #1 \right)}
\newcommand{\lb}[1]{\left[ #1 \right]}
\newcommand{\lc}[1]{\left\{ #1 \right\}}
\newcommand{\boverc}{\frac{b}{c}}

% \graphicspath{{figs/}}

\maketitle

\begin{abstract}
  Indirect reciprocity is a key explanation for the exceptional magnitude of cooperation among humans.
  This literature suggests that a large proportion of human cooperation is driven by social norms and individuals' incentives to maintain a good reputation.  
  This intuition has been formalized with two types of models. 
  In public assessment models, all community members are assumed to agree on each others' reputations; in private assessment models, people may have disagreements. 
  Both types of models aim to understand the interplay of social norms and cooperation. 
  Yet their results can be vastly different. 
  Public assessment models argue that cooperation can evolve easily, and that the most effective norms tend to be stern. 
  Private assessment models often find cooperation to be unstable, and successful norms show some leniency. 
  Here, we propose a model that can organize these differing results within a single framework. 
  We show that the stability of cooperation depends on a single quantity:  the extent to which individual opinions turn out to be correlated.
  This correlation is determined by a group's norms and the structure of social interactions. 
  In particular, we prove that no cooperative norm is evolutionarily stable when individual opinions are statistically independent. 
  These results have important implications for our understanding of cooperation, conformity, and polarization. 
\end{abstract}

%%%%%%%%%%%%%
%% INTRODUCTION %%
%%%%%%%%%%%%%

\section{Introduction}

Indirect reciprocity can explain why unrelated individuals -- even complete strangers -- might cooperate with each other~\cite{melis:ptrs:2010,rand:TCS:2013,nowak2006five}.
This explanation suggests that people cooperate because they wish to maintain a positive reputation within their community. 
There is a number of empirical patterns in line with this view. 
For example, humans act more pro-socially when their actions are widely observable~\cite{semmann:BES:2004,Engelmann:GEB:2009,yoeli:PNAS:2013}; 
they seek information to gauge the social standing of their interaction partners~\cite{Swakman:EHB:2016,Molleman:PRSB:2016};
and they are more likely to help those with a positive reputation~\cite{wedekind:Science:2000,wedekind2002long,milinski:Nature:2002}. 

To better understand these empirical patterns, theoretical studies work with two types of models. 
The first type, the public assessment model~\cite{ohtsuki2004should,ohtsuki2006leading,pacheco2006stern,santos2016social,santos2018social,ohtsuki2007global,masuda2012ingroup,berger2011learning,nakamura2014indirect,ohtsuki2015reputation,clark2020indirect,murase2022social,murase2023indirect}, assumes that all community members agree on each other's reputations.
In particular, if one member thinks highly of some third party, then so does everyone else. 
Such an assumption may appear as rather extreme. 
Yet it has been hugely successful, mostly because it drastically simplifies a model's mathematical complexity. 
Based on this assumption, Ohtsuki and Iwasa~\cite{ohtsuki2004should,ohtsuki2006leading} were able to identify eight social norms that can stabilize cooperation.
These norms, known as the `leading eight', have been widely studied since, even though there are other evolutionarily stable norms that equally support cooperation~\cite{murase2023indirect}.

\begin{figure*}
  \centering
  \includegraphics[width=0.92\textwidth]{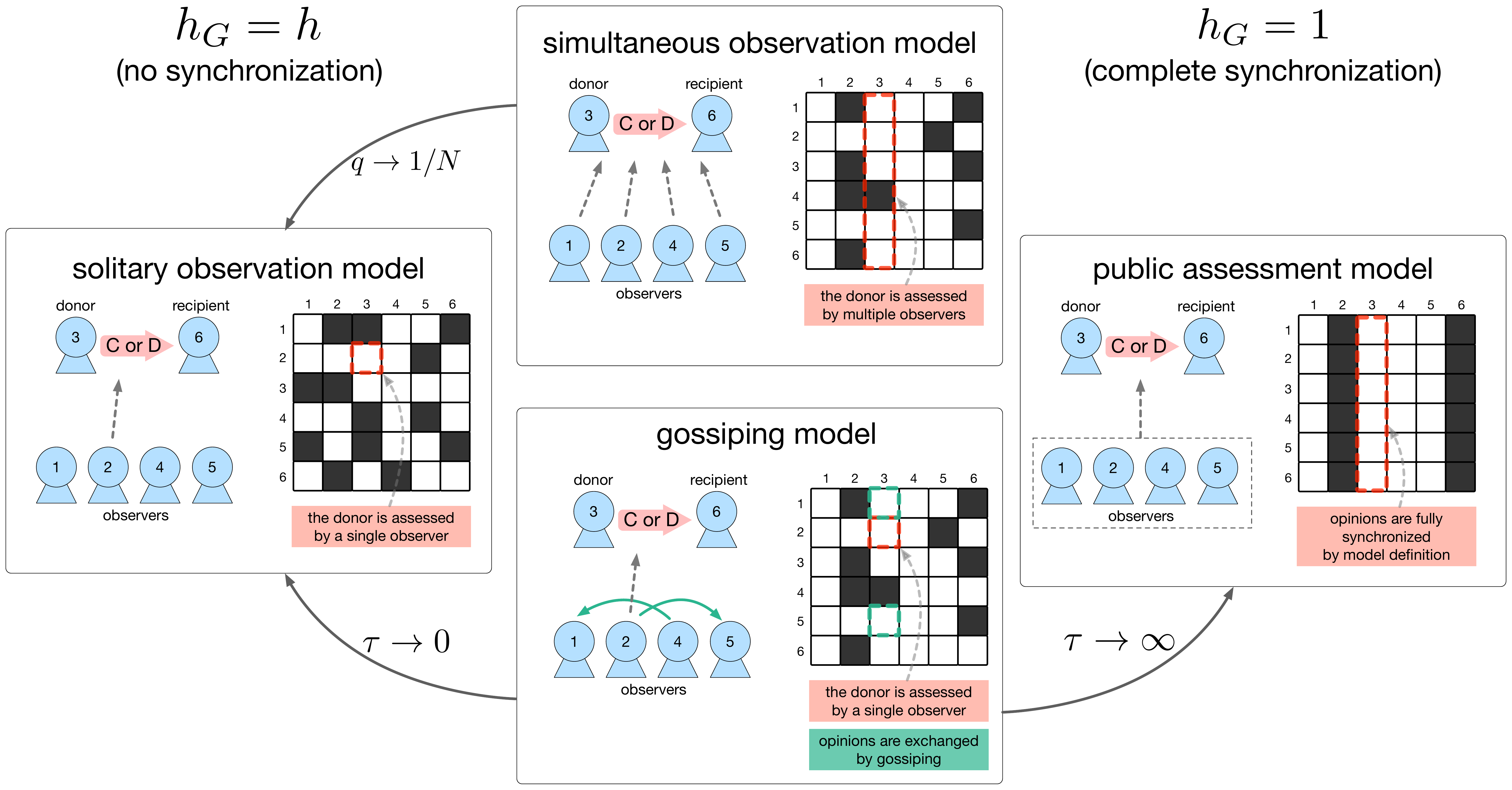}~\\[-0.2cm]
  \caption{
    \textbf{Schematic illustration of the models considered in this paper}.
    We consider a population of $N$ players (here, $N\!=\!6$).
    At each time step, randomly chosen donor-recipient pairs play the donation game.
    In each game, the donor decides whether to cooperate ($C$) or to defect ($D$) according to its action rule $P$.
    Other players may observe the donor's decision and update its reputation according to their assessment rule $R$.
    The resulting reputations can be represented by an image matrix. 
    It records how each player assesses every other at a given time. 
    Here, we represent the image matrix by a square with six rows and six columns. 
    Black entries indicate that the respective row player thinks negatively of the column player. 
    White entries indicate positive opinions. 
    In the following, we revisit four classical models that differ in how these image matrices are updated. 
    On the left, there is the solitary observation model. 
    Here, each interaction is observed by a single player.
    As a result, the rows of the image matrix turn out to be independent.
    In the top-middle, there is the simultaneous observation model, where each action is observed by any given player with probability $q$.
    As the observation probability $q$ approaches $1/N$, the model simplifies to the solitary observation model.
    In the bottom-middle, there is the gossiping model by Kawakatsu et al~\cite{kawakatsu2024mechanistic}.
    In this model, players share their opinions with other population members. 
    When the gossip duration $\tau$ approaches zero, the model becomes equivalent to the solitary observation model.
    When $\tau \to \infty$, the model approaches the public assessment model depicted on the right.
    In that model, opinions are perfectly synchronized.
    }
  \label{fig:schematic_diagram}
\end{figure*}

The second type, the private assessment model~\cite{uchida2010effect,uchida2013effect,perret2021evolution,hilbe2018indirect,schmid2021evolution,schmid2021unified,schmid2023quantitative,fujimoto2022reputation,fujimoto2023evolutionary,fujimoto2024leader,lee2021local,okada2017tolerant,okada2018solution,okada2020two,kawakatsu2024mechanistic,kessinger2023evolution,radzvilavicius2019evolution,radzvilavicius2021adherence,krellner2021pleasing}, recognizes that individuals may differ in how they view others.
A mathematical analysis of this type of model is more complex.  
Private assessment models need to keep track of how each population member thinks of everyone else. 
The situation can be represented by an `image matrix', see Fig.~\ref{fig:schematic_diagram}.
Each row of this matrix represents an individual who evaluates the reputations of other group members. 
Each column represents whose reputation is evaluated. 
The entries of this matrix correspond to the assigned reputations (in Fig.~\ref{fig:schematic_diagram}, they are black or white, i.e. `bad' or `good'). 
This image matrix can change in time, depending on whether individuals cooperate, how observable their actions are, and on the social norm in place. 
With respect to the observability of individual actions, one can further distinguish three types of models: simultaneous observation models~\cite{hilbe2018indirect,schmid2021evolution,schmid2021unified,schmid2023quantitative,fujimoto2022reputation,fujimoto2023evolutionary,fujimoto2024leader,lee2021local}, solitary observation models~\cite{okada2017tolerant,okada2018solution,okada2020two}, and models incorporating communication~\cite{kawakatsu2024mechanistic,kessinger2023evolution,radzvilavicius2019evolution,radzvilavicius2021adherence,krellner2021pleasing}.

Each of these model types is well-established. 
However, they have typically been studied in isolation. 
Moreover, the various private assessment and public assessment models often yield conflicting findings.
For instance, public assessment models often find that a particular leading-eight norm, `Stern Judging', is most favorable for the evolution of cooperation~\cite{pacheco2006stern,santos2018social}. 
In contrast, in most private assessment models, the very same social norm proves to be highly inefficient~\cite{uchida2010effect,hilbe2018indirect}.
These discrepancies make it difficult to assess whether indirect reciprocity can sustain cooperation at all, and which social norms are most effective. 

Here, we propose a general framework to understand the literature through the lens of opinion synchronization.
Our framework contains the previous models as special cases, and it systematically reproduces their results.
The two key quantities in our model are the variables $h$ and $h_G$. 
The first variable $h$ describes how often, on average, individuals assess others as good. 
The second variable $h_G$ captures to which extent opinions are synchronized. 
For example, consider three distinct individuals, Alice, Bob, and Charlie.
Then $h_G$ corresponds to the conditional probability that Charlie views Bob as good, given that Alice does.
The solitary observation model~\cite{okada2017tolerant,okada2018solution,okada2020two} corresponds to the case $h_G\!=\!h$. 
Here, opinions are statistically independent. 
At the other extreme, the public assessment model assumes $h_G\!=\!1$.  
Here, opinions are perfectly correlated. 
The simultaneous observation model and models that allow for communication are in between these two extremes. 
They satisfy $h\!\leq\!h_G\!\leq\!1$ (Fig.~\ref{fig:schematic_diagram}). 
One important finding is that stable cooperation requires opinions to be sufficiently synchronized.
In particular, when opinions are statistically independent, cooperative social norms are either unstable or at most neutrally stable. 
As opinions become more synchronized, say because of shared experiences or gossip, cooperation can be established more easily. 

Our results make an interesting connection to the literature on conformity~\cite{cialdini2004social,KrebsDenton:PNAS:2020,KaledaDenton:PNAS:2022}.
In the context of evolutionary game theory, this literature often argues that conformity may enhance the chance people cooperate, even though cooperators are at a slight disadvantage~\cite{henrich:JTB:2001,Romano:PsySci:2017}. 
Our model offers a slightly different perspective. 
In models of indirect reciprocity, a certain kind of conformity is what allows cooperation to be in everybody's interest. 
To this end, we interpret conformity as the degree to which individual opinions (about others) are synchronized.
This synchronicity is an emerging trait, which depends on the social norm in place and on the structure of social interactions. 
In particular, it depends on how publicly observable interactions are, and to which extent individuals exchange gossip. 
We find that only when opinions are sufficiently synchronized, the mechanism of indirect reciprocity can be effective.

%%%%%%%%%
%%  MODEL  %%
%%%%%%%%%

\section*{Model}

We consider a  model of indirect reciprocity that interpolates between public and private assessment models.
There is a population of size $N \!\gg\! 1$. 
The members of this population (referred to as `players') engage in a sequence of donation games.
Each round, one player is randomly selected to act as a donor. 
Another player is randomly selected to be the recipient.
Donors can either cooperate~($C$) or defect~($D$).
A cooperating donor pays a cost $c$ to provide a benefit $b\!>\!c$ to the recipient.
A defecting donor pays no cost and provides no benefit. 
This elementary process is repeated for many rounds, with changing donors and recipients. 
Over the course of these games, the players accumulate their payoffs.

During this sequence of the donation games, players form opinions about each other.
The opinion player $i$ holds about $j$ is denoted as $m_{ij}$. 
The corresponding matrix $M\!=\!(m_{ij})$ is referred to as an image matrix.
Following the convention of the previous literature, we assume opinions are binary, either good ($G$) or bad ($B$). 
These opinions can change over time.
Moreover, one player's opinion about another player is not necessarily shared by all other population members. 
In terms of the image matrix, this means that the entries in any given column do not need to be the same.

A strategy of player $i$ is a combination $(P_i, R_i)$ of an action rule and an assessment rule  (in line with the indirect reciprocity literature, we use the terms `strategy' and `social norm' synonymously).
Herein, we consider stochastic second-order strategies~\cite{murase2023indirect}.
The action rule $P_i(m_{ij})$ determines the cooperation probability of donor $i$, given the donor's opinion $m_{ij}$ about recipient~$j$.
Let us denote the realized action as $A_{ij} \in \{C, D\}$.
An observer $k$ assesses the donor $i$ based on the donor's action $A_{ij}$ and the observer's opinion about the recipient $m_{kj}$.
The donor is assessed as good with probability $R_k(m_{kj}, A_{ij})$. 
Otherwise the donor is assessed as bad.
Table~\ref{tab:models} gives a few examples of well-known assessment rules. 
Note that because neither the action rule nor the assessment rule depend on a player's self image, the diagonal entries $m_{ii}$ of the image matrix are irrelevant. 

Assessments can be subject to errors.
With probability~$\mu_a$, observers assign the opposite reputation to a donor, compared to the assignment prescribed by the assessment rule. 
As a result, the effective assessment rule becomes
\begin{equation}
  \tilde{R}\lp{A, X} = \lp{1-\mu_a}R\lp{A, X} + \mu_a \lb{1-R\lp{A, X} }.
\end{equation}
In the following, we refer to $\tilde{R}$ as $R$ for simplicity.
Moreover, we limit ourselves to the case $0 \!<\! \mu_a \!\le\! 1/2$ (if the error probability was greater than 1/2, the meaning of `good' and `bad' would be flipped). 
Because the error probability $\mu_a$ is positive, effective assessments are stochastic, $0\! <\! R\! <\! 1$.
This stochasticity ensures that the reputation dynamics is ergodic. 
As a result, the time average of the image matrix reaches a unique stationary state that is independent of the initial conditions (for details, see \SI).

Based on this general framework, we consider four different models to update the image matrix.
First, we consider the solitary observation model~\cite{okada2017tolerant,okada2018solution,okada2020two}, as depicted on the left hand side of Fig.~\ref{fig:schematic_diagram}.
At each donation game, a single observer is randomly chosen to update their opinion about the donor.
Since only a single element of the image matrix is updated each round, the elements in each column are statistically independent.
This simplification allows for a fully analytical treatment.
Here, we only need to keep track of the fraction $h$ of good entries in the image matrix. 

Second, we consider the simultaneous observation model, depicted in the top of Fig.~\ref{fig:schematic_diagram}.
This model allows each population member to observe the donor's action with some fixed observation probability $q$.
In particular, several observers may witness the same interaction simultaneously. 
As $q$ becomes small (compared to the population size), the model becomes equivalent to the solitary observation model.
However, for general $q$, an analytical treatment is more challenging. 
Except for a few special cases, this model often requires numerical simulations. 

Third, we consider the gossiping model by Kawakatsu et al.~\cite{kawakatsu2024mechanistic}, as depicted in the bottom of Fig.~\ref{fig:schematic_diagram}.
This model allows for communication among the players.
Donation games are played between all pairs in the population.
Each player updates its opinion about each of the other players based on their action as the donor in a randomly chosen game.
Thus, on average, each action is observed by a single observer, as in the solitary observation model.
The assessment phase is followed by a gossip phase, where players exchange opinions.
This phase consists of several gossiping events.
During each event, a randomly chosen pair of players exchange their opinions about a randomly chosen population member. 
As a result, a randomly chosen entry of the image matrix replaces another randomly chosen entry in the same column.
The gossiping events occur repeatedly for a certain number of times, characterized by the gossip duration~$\tau$.
The parameter $\tau$ controls the degree of synchronization of opinions.
When $\tau \!=\! 0$, the model is equivalent to the solitary observation model. 
As  $\tau \to \infty$, we recover the public assessment model.

The public assessment model is depicted on the right hand side of Fig.~\ref{fig:schematic_diagram}.
This model assumes that all players always have the same opinion about each co-player. 
This assumption implies that all entries in any given column of the image matrix are the same. 
The public assessment model represents the most extreme case of opinion synchronization.
Because it allows for a comfortable analytical treatment, it is often used as a benchmark.

\begin{table}
  \centering
  \begin{tabular}{c|cccc}
    Recipient's reputation & \multicolumn{2}{c}{$G$} & \multicolumn{2}{c}{$B$} \\
    Donor's action         & $C$ & $D$ & $C$ & $D$ \\
    \hline
    Simple Standing (L3)   & $1$ & $0$ & $1$ & $1$ \\
    Stern Judging (L6)     & $1$ & $0$ & $0$ & $1$ \\
    Image Scoring          & $1$ & $0$ & $1$ & $0$ \\
    Shunning               & $1$ & $0$ & $0$ & $0$ \\
    \hline
  \end{tabular}
  ~\\[0.2cm]
  \caption{
    Examples of the assessment rules.
    {\normalfont We represent four well-known assessment rules. In each case, the probability $R(X, A)$ of assessing the donor as good is shown for each combination of the donor's action $A$ and the recipient's reputation $X$.
    Among these four rules, only the leading-eight strategies L3 and L6 promote evolutionarily stable cooperation in the public assessment model.}
    }
  \label{tab:models}
\end{table}

\section*{Analysis of the solitary observation model}

To illustrate our approach, we first analyze the solitary observation model. 
Here, opinions of different individuals turn out to be statistically independent, which simplifies the analysis. 
In subsequent sections, we generalize this approach to allow for arbitrary correlations between opinions. 

For the solitary observation model, we show that evolutionarily stable cooperation is impossible.
More specifically, we show that for any second-order resident strategy, either unconditional cooperation (ALLC) or unconditional defection (ALLD) is always a best response.
In the main text, we present an outline of the analysis; all details are in the  {\SI}.

We consider a monomorphic resident population in which everyone uses the strategy $(P, R)$.
Because of the assumption of solitary observations, the entries of the image matrix are statistically independent. 
Moreover, due to the ergodicity of the process, the fraction $h$ of good entries in the image matrix converges to a unique stationary value after a sufficiently long time. 
This value satisfies the equation
\begin{equation}
  \begin{split}
  h &= h^2 R_P\lp{G, G} + h(1-h) \lb{ R_P\lp{G, B} + R_P\lp{B, G} }  \\
    &\quad + \lp{1-h}^2 R_P\lp{B, B}.
  \end{split}
  \label{eq:h_ast_solitary}
\end{equation}
Here, the variable
\begin{equation}
R_P(X, Y) \equiv P(X)R(Y,C) + \lb{ 1-P\lp{X} } R(Y,D).
\label{eq:R_P}
\end{equation}
denotes the probability that an observer $k$ assesses the donor $i$ as good, given their initial opinions about recipient~$j$,  $m_{ij} \!=\! X$ and $m_{kj} \!=\! Y$.
\eqref{eq:h_ast_solitary} provides an implicit formula for the average fraction $h$ of well-assessed individuals in a homogeneous resident population. 

Next, we consider a small minority of players who deviate towards a different strategy.
For brevity, we refer to the deviating players as mutants. 
To compute whether mutants can invade, we need to compute their payoffs, which in turn depends on how often residents cooperate with a mutant. 
To do this computation, it turns out to be useful to consider the average cooperation rate of a mutant towards the residents, $p_{\rm mut \to res}$.
Importantly, here we do not have to define the specific form of the mutant's action and assessment rule. 
For the following calculations, only the mutants' actions matter, irrespective of how complex their underlying strategies are. 
Let $H$ denote the average probability that a resident considers a mutant to be good. 
After a sufficiently long time, $H$ converges to a unique stationary fixed point, defined by
\begin{equation}
  \begin{split}
  H &= p_{\rm mut \to res} \lb{ h R\lp{G,C} + \lp{1 - h} R\lp{B,C} }  \\
          &\quad + \lp{1 - p_{\rm mut \to res}} \lb{ h R\lp{G,D} + \lp{1 - h} R\lp{B,D} }.  \\
  \end{split}
\end{equation}
In particular, we obtain the following formula for how likely residents are to cooperate with the mutant, 
\begin{equation}
\begin{split}
  p_{\rm res \to mut} &= H P\lp{G} + \lp{1\!-\!H} P\lp{B}   \\
    &= p_{\rm mut \to res} P_b + P_0.
\end{split}
\label{eq:pc_res_mut}
\end{equation}
The coefficient $P_0$ is independent of the mutant's strategy; its exact form is given in the \SI. 
The other coefficient $P_b$ can be interpreted as the expected net reward for cooperation, 
\begin{equation}
  \begin{split}
  P_{b} &\equiv \lb{P\lp{G} - P\lp{B} } \left\{ h \lb{R\lp{G,C} - R\lp{G,D} } \right.  \\
        &\qquad \left. + \lp{1-h} \lb{ R\lp{B,C} - R\lp{B,D} } \right\}.
  \end{split}
  \label{eq:P_b}
\end{equation}
The factor $\lb{ P\lp{G} \!-\! P\lp{B} }$ reflects how valuable a good reputation is.
The larger this factor, the more likely a good player receives cooperation compared to a bad player.
The second factor $\lc{ h\lb{ R\lp{G,C} \!-\! R\lp{G,D} } + \lp{1\!-\!h} \lb{ R\lp{B,C} \!-\! R\lp{B,D} } }$ indicates how much more likely the mutant gets a good reputation by cooperating.

\begin{figure}
  \centering
  \includegraphics[width=0.48\textwidth]{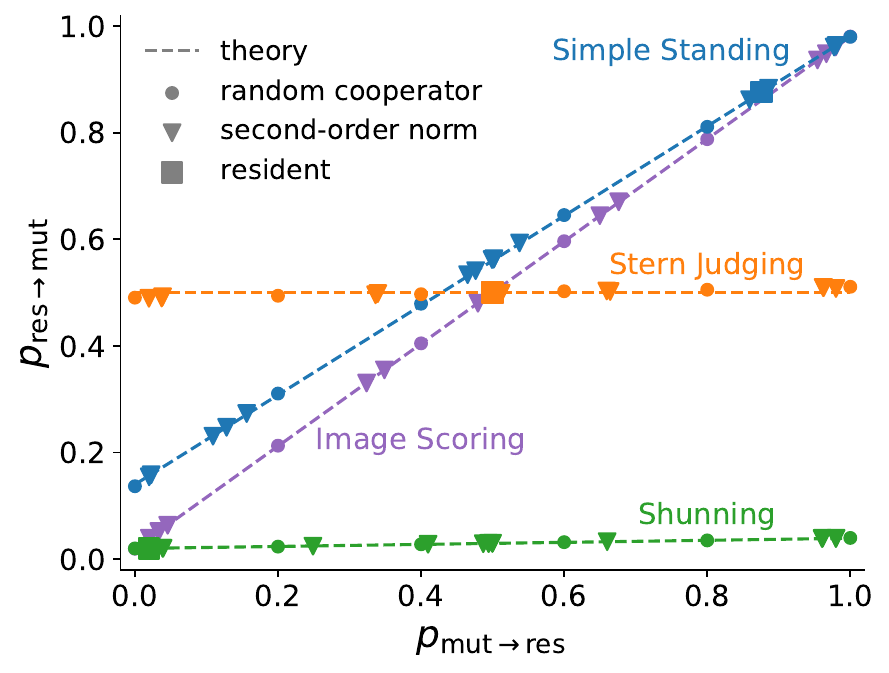}
  \caption{
    \textbf{Relationship of the cooperation levels between residents and mutants.}
    As resident norms, we consider Simple Standing, Stern Judging, Image Scoring, and Shunning, as defined in Table~\ref{tab:models}.
    Dashed lines are theoretical predictions obtained from Eq.~(\ref{eq:pc_res_mut}). 
    Points are obtained from numerical simulations for $N\!=\!100$ and $\mu_a\!=\!0.02$ (see {\meth}~A for details).
    Circles indicate results for unconditionally cooperating mutants with cooperation probability $p_{\rm mut \to res} \in \{0, 0.2, 0.4, 0.6, 0.8, 1.0\}$, respectively.
    Triangles are the results for deterministic second-order mutants.
    We observe that regardless of the mutant strategy, a resident's average cooperation rate towards the mutant, $p_{\rm res \to mut}$, is a linear function of the mutant's cooperation rate $p_{\rm mut \to res}$.
    }
  \label{fig:pc_mut_res}
\end{figure}

Crucially, \eqref{eq:pc_res_mut} indicates that $p_{\rm res \to mut}$ is a linear function of $p_{\rm mut \to res}$.
In Fig.~\ref{fig:pc_mut_res}, we illustrate this linear relationship with numerical simulations. 
For these simulations, we consider four different resident strategies (the same as in Table~\ref{tab:models}). 
The resident strategy is adopted by $N\!-\!1$ population members. 
The remaining mutant player either cooperates unconditionally with a certain probability, or adopts a deterministic second-order strategy. 
Given this population composition, we simulate the game dynamics described in the Model section. 
Over the course of the simulation, we record how often the mutant cooperates with the residents, and conversely how often residents cooperate with the mutant. 
As predicted by \eqref{eq:pc_res_mut}, we find a perfect linear relationship between these cooperation rates. 
This relationship is independent of the complexity of the mutant strategy. 

Because cooperation rates obey a linear relationship, also the mutant's payoff can be written as a linear function,
\begin{equation}
\label{eq:mutant_payoff}
  \begin{split}
  \pi_{\rm mut} &= bp_{\rm res \to mut} - cp_{\rm mut \to res}   \\
                &= \lp{b P_b - c} p_{\rm mut \to res} + bP_0.
  \end{split}
\end{equation}
This representation of the mutant's payoff is useful because linear functions are comparably easy to analyze.
In particular, they typically attain a unique maximum on the boundary of the domain (in this case, for $p_{\rm mut \to res}\!\in\!\{0,1\}$). 
For \eqref{eq:mutant_payoff}, we conclude that the mutant maximizes its payoff when
\begin{equation}
  p_{\rm mut \to res} = \begin{cases}
    1 & {\rm when \quad} b P_{b} > c  \\
    {\rm any} & {\rm when \quad} b P_{b} = c \\
    0 & {\rm when \quad} b P_{b} < c
  \end{cases},
  \label{eq:optimal_mutant}
\end{equation}
Thus, a mutant can always maximize its payoff by playing ALLC or ALLD.
In contrast, conditional cooperation is generally not optimal. 
The only exception arises when $b P_{b} \!=\! c$, which corresponds to the Generous Scoring norm~\cite{schmid2021unified}.
But even here, any mutant strategy obtains the same payoff as the residents; thus any mutant can invade by neutral drift.
In this sense, any conditionally cooperative strategy is unstable. 

For the above result, the assumption of statistically independent opinions is crucial. 
For an intuition, consider the best action for the donor Alice, towards the recipient Bob, when observed by Charlie.
If Alice wants to maximize her long-term payoff, and if opinions are statistically independent, Alice does not need to change her actions depending on her own opinion of Bob.
For her, only the opinion of Charlie, who is a potential future interaction partner, matters. 
It would be best for Alice if she could condition her action based on Charlie's opinion. 
However, because of the independence assumption, her own view of Bob provides Alice with no information about Charlie's opinion. 
Conversely, from Charlie's perspective, Alice looks as if she is randomly cooperating with a certain probability, even as Alice changes her actions depending on her own opinion.
Thus, if the long-term benefit of cooperation exceeds the immediate cooperation costs, $b\,P_b \!>\! c$, Alice should always cooperate.
If the long-term benefit is below this threshold, Alice should always defect. 
In the \SI, we derive an analogous conclusion for second-order norms with non-binary reputations.

The above results assume mutants to be infinitesimally rare.
A similar argument, however, also applies when the fraction of mutants is strictly positive: 
no resident strategy can prevent the neutral invasion by randomly cooperating mutants.
To see why, let $p_r$ be the cooperation level of the pure resident population.
From a resident's viewpoint, other residents act as if they are randomly cooperating with probability $p_{r}$.
Thus, a resident cannot distinguish another resident from a mutant who unconditionally cooperates with probability $p_{r}$. 
This neutrality remains even after the share of mutants in the population increases.
Hence, the residents are subject to neutral invasion for any mixed population of residents and mutants, as shown in Fig.~S2.
This conclusion only relies on the  statistical independence of opinions. 
In particular, the same conclusion applies to more complex strategies, such as strategies with non-binary opinions~\cite{murase2022social,schmid2023quantitative,lee2021local}, higher-order strategies~\cite{santos2018social}, and strategies with dual-reputation updates~\cite{murase2023indirect}.

The above arguments, however, require that there are at most two strategies in the population.
Hence, this result does not rule out the possibility that a mixture of several strategies forms stable cooperation.
For instance, previous research suggests that  Simple Standing (L3) and ALLC can stably coexist when these strategies compete with ALLD~\cite{okada2017tolerant,okada2018solution,okada2020two,radzvilavicius2019evolution,radzvilavicius2021adherence,kessinger2023evolution}.
We do not have a general argument on the stability of such a mixture.
It is unknown whether the mixture is stable against a more diverse set of strategies, or invadable by a certain mutant such as random cooperators.

\section*{Analysis on models with correlated opinions}

In a next step, we generalize the above approach to allow for correlated opinions. 
In that case, we show there are evolutionarily stable norms that sustain cooperation. 
Again, in the following, we present an outline of the analysis; the {\SI} contains all details.

Consider three players, Alice, Bob, and Charlie, whose opinions may be correlated. 
The first player, Charlie, considers Bob as good with probability $h$.
Let $h_G$ denote the conditional probability that Alice assigns a good reputation to Bob as well, given that Charlie does. 
Similarly, let $h_B$ denote the conditional probability that Alice assigns a good reputation to Bob given that Charlie does not. 
The three probabilities $h$, $h_G$, and $h_B$ are not independent of each other.
After all, the probability that two randomly chosen players have opposite opinions is equal to both $h(1\!-\!h_G)$ and $(1\!-\!h)h_B$.
Therefore, $h(1\!-\!h_G) = (1\!-\!h)h_B$ needs to hold.
In particular, when $h_G\! =\! h$, $h_B \!=\! h$ as well. 
This corresponds to the special case of independent opinions.
If Alice's opinion is positively correlated with Charlie's, then $h_G \!>\! h \!>\! h_B$; Alice is more likely to consider Bob as good when Charlie does.
In the perfectly correlated case (i.e., for the public assessment model), $h_G \!=\! 1$ and $h_B \!=\! 0$.

At this moment, we do not make any further assumptions on how exactly opinions get correlated.
For example, Alice and Charlie could have both witnessed Bob's behavior simultaneously~\cite{hilbe2018indirect,schmid2021evolution,schmid2023quantitative,fujimoto2022reputation,fujimoto2023evolutionary,fujimoto2024leader,krellner2023we}. 
Alternatively, they could have exchanged opinions by gossiping~\cite{kessinger2023evolution,kawakatsu2024mechanistic,Pan:PNAS:2024}, or obtained relevant information from some public institution~\cite{radzvilavicius2021adherence}.
Our results are independent of how correlations are achieved.

Consider a monomorphic population in which all players use the norm $\{P, R\}$.
As before, we can derive an equation that needs to be satisfied in the stationary state,
\begin{equation}
  \begin{split}
  h &= h h_G            R_P\lp{G, G}  \\
    &\quad + h\lp{1-h_G}      \lb{ R_P\lp{G, B} + R_P\lp{B, G} }  \\
    &\quad + \lp{1-h}\lp{1 - h_B} R_P\lp{B, B}.
  \end{split}
  \label{eq:h_ast_correlated}
\end{equation}
In particular, there is only one degree of freedom:
given the strategy $\{P, R\}$, $h$ is determined once $h_G$ is fixed (and vice versa).
The exact value of $h_G$ depends on the considered model.
Once we define a model (such as solitary observation, or the simultaneous observation model), all three quantities $h$, $h_G$, $h_B$ are uniquely determined.
For the solitary observation model, $h \!=\! h_G \!=\! h_B$, and Eq.~(\ref{eq:h_ast_correlated}) simplifies to Eq.~(\ref{eq:h_ast_solitary}).
For other models, it may not be possible to obtain analytical expressions for $h$ and $h_G$. 
In that case, we need simulations; see  {\meth} for details.

Next, we consider a mutant having a different action rule $P'$ but the same assessment rule $R$.
We are going to show that the optimal action rule is either ALLC, ALLD, or conditional cooperation with $P'(G) \!=\! 1$ and $P'(B) \!=\! 0$.
We derive under which conditions conditional cooperation is optimal for given $\{P, R\}$, $h$, and $h_G$.
To this end, let $H$ be the average probability that a mutant is assessed as good. 
Again, $H$ converges to a unique stationary value after a sufficiently long time. 
Since $H$ can be written as a function of $h$ and $h_G$, the probability that a resident cooperates with a mutant becomes
\begin{equation}
  \begin{split}
  p_{\rm res \to mut} &= H P(G) + \lp{ 1-H } P(B)   \\
                      &= \lb{ hh_G \Delta_G + h\lp{ 1-h_G } \Delta_B } P'(G)   \\
                      &+ \lb{ h\lp{1\!-\!h_G} \Delta_G + \lp{1\!-\!2h\!+\!hh_G} \Delta_B } P'(B)  \\
                      &+ P_1.
  \end{split}
\end{equation}
Here, $\Delta_G$ and $\Delta_B$ are defined as
\begin{equation}
\begin{split}
  \Delta_G &\equiv \lb{ R\lp{G, C} - R\lp{G, D} } \lb{ P\lp{G} - P\lp{B} }  \\
  \Delta_B &\equiv \lb{ R\lp{B, C} - R\lp{B, D} } \lb{ P\lp{G} - P\lp{B} }.
\end{split}
\end{equation}
Moreover, $P_1$ is a constant term that does not depend on $P'$.
Similarly, the probability that a mutant cooperates with a resident is
\begin{equation}
  p_{\rm mut \to res} = h P'(G) + (1-h) P'(B)
\end{equation}
Therefore, the mutant's payoff is
\begin{equation}
\begin{split}
  \pi_{\rm mut} &= b p_{\rm res \to mut} - c p_{\rm mut \to res}    \\
                &= \alpha_G P'(G) + \alpha_B P'(B) + b P_1,
\end{split}
\end{equation}
where we defined
\begin{equation}
\begin{split}
  \alpha_G &\equiv h h_G ( b\Delta_G \!-\! c ) + h(1\!-\!h_G) ( b\Delta_B \!-\! c )  \\
  \alpha_B &\equiv h(1\!-\!h_G) ( b\Delta_G \!-\! c ) + (1\!-\!2h\!+\!hh_G) ( b\Delta_B \!-\! c ).
\end{split}
\label{eq:alpha_G_B_definition}
\end{equation}
Since $\pi_{\rm mut}$ is a linear function of $P'(G)$ and $P'(B)$, the best action rule $\hat{P'}$ for the mutant is summarized as follows:
\begin{equation}
\begin{split}
  \hat{P'}(G) &= \begin{cases}
    1  &  \text{when $\alpha_G > 0$}  \\
    \text{any} & \text{when $\alpha_G = 0$}  \\
    0  &  \text{when $\alpha_G < 0$}
    \end{cases}  \\
  \hat{P'}(B) &= \begin{cases}
    1  &  \text{when $\alpha_B > 0$}  \\
    \text{any} & \text{when $\alpha_B = 0$}  \\
    0  &  \text{when $\alpha_B < 0$}
    \end{cases}.  \\
\end{split}
\label{eq:mut_best_action}
\end{equation}
From this equation, we conclude that the best action rule $\hat{P'}$ needs to be deterministic, except for the special cases of $\alpha_G \!=\! 0$ or $\alpha_B \!=\! 0$.
When $\alpha_G$ and $\alpha_B$ have the same sign, unconditional cooperation or defection is the best action.
Conditional cooperation is optimal when $\alpha_G \!>\! 0$ and $\alpha_B \!<\! 0$.
In that case, the average cooperation rate of the population coincides with the fraction $h$ of individuals with a good reputation. 
Note that when $h\!=\!h_G$, $\alpha_G$ and $\alpha_B$ have the same sign, reproducing the conclusion of the previous section.

\begin{figure*}[t]
\centering
\includegraphics[width=0.75\textwidth]{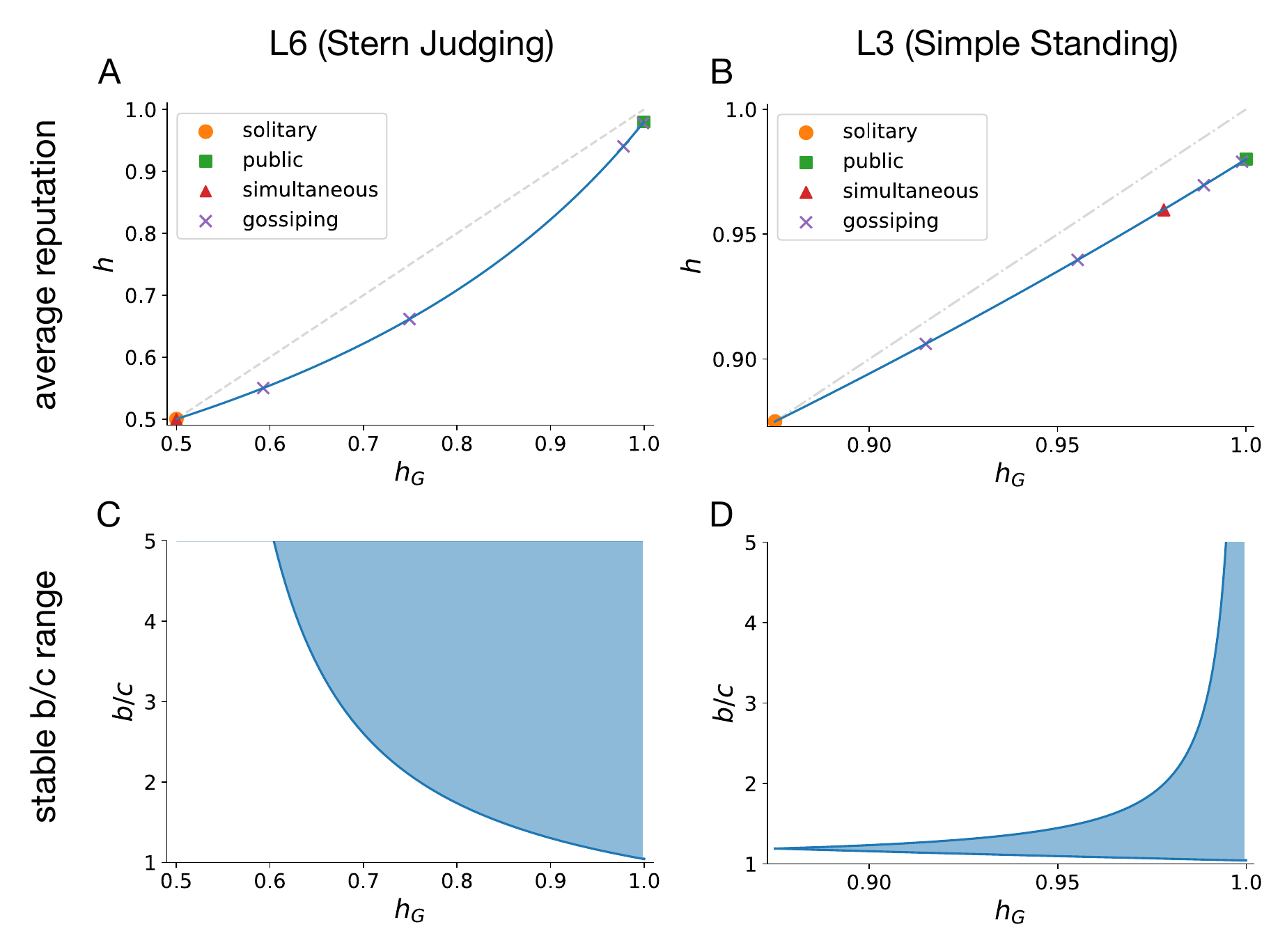}
\caption{
  \textbf{(A,B)} The relationship between $h$ and $h_G$ for the norms L6 (Stern Judging) and L3 (Simple Standing).
  The gray dashed lines indicates the case $h\!=\!h_G$, which is shown as a reference.
  For both norms, $h$ increases as $h_G$ increases.
  The leftmost $h_G$ is the minimal value obtained for the solitary observation model; here, $h_G\!=\!h$.
  The rightmost $h_G$ corresponds to the public assessment model $h_G \!=\! 1$.
  Results for the simultaneous observation model with $N \!=\! 100$ and $q \!=\! 1$ are shown as red triangles.
  Purple crosses indicate the gossiping model with $\tau \!=\! 0.1$, $0.3$, $1$, and $3$, from left to right.
  \textbf{(C, D)} The blue area indicates the benefit-to-cost ratios $b/c$ for which a conditionally cooperative action rule is stable.
  As $h_G$ increases, the stable range of $b/c$ expands.
  We use an assessment error rate is $\mu_a \!=\! 0.02$.
  }
\label{fig:stable_bc_range}
\end{figure*}

In Fig.~\ref{fig:stable_bc_range}, we illustrate these results for the norms Stern Judging (L6) and Simple Standing (L3).
The top panels, Fig.~\ref{fig:stable_bc_range}A and \ref{fig:stable_bc_range}B, show the functional relationship between $h$ and $h_G$.
In each case, $h$ increases as $h_G$ increases: the cooperation level rises as opinions are more synchronized (such a positive relationship does not need to hold for other norms). 
In addition, we depict the realized values of $h$ and $h_G$ for each of the different model types considered. 
The most extreme models are  the solitary observation model (to the left) and the public assessment model (to the right), respectively.
The bottom panels, Fig.~\ref{fig:stable_bc_range}C and \ref{fig:stable_bc_range}D, show the ranges of the benefit-to-cost ratio $b/c$ for which conditional cooperation is stable, calculated from Eq.~(\ref{eq:mut_best_action}).
The stable ranges of $b/c$ expand with $h_G$, indicating that cooperation is easier to sustain when opinions are more synchronized.
By superimposing the upper and the lower panels, we can also infer for which social structure each of the two norms is stable. 
  For example, for Stern Judging, the upper panel suggests that in the simultaneous observation model, we obtain $h_G\!=\!0.5$. 
  For this value of $h_G$, the lower panel suggests that Stern Judging is unstable, for any benefit-to-cost ratio, in line with previous research~\citep{uchida2013effect}. 
  In contrast, for Simple Standing, simultaneous observations lead to $h_G\!\approx\!0.98$, which permits stable cooperation if $b/c$ is sufficiently small.

\section*{Application to specific models}

In the following, we apply this general formalism to three special cases: the public assessment model, the simultaneous observation model, and the gossiping model.
In this way, we show that we are able to systematically reproduce a large set of previous results within a single framework.

\subsection*{Public assessment model}

First, we consider public assessment. 
Here, opinions are perfectly synchronized, $h_G \!=\! 1$ and $h_B \!=\! 0$.
For this model, the set of stable norms that sustain cooperation has been fully characterized in Ref.~\cite{murase2023indirect}.
According to Eq.~(\ref{eq:mut_best_action}), conditional cooperation with $P(G)\!=\!1$ and $P(B)\!=\!0$ is optimal when
\begin{equation}
\begin{cases}
  \alpha_G > 0 &\iff b \Delta_G > c   \\
  \alpha_B < 0 &\iff b \Delta_B < c
\end{cases}.
\end{equation}
A conditionally cooperative response is the unique best action rule when
\begin{equation}
\begin{cases}
  b \lb{ R\lp{G, C} - R\lp{G, D} } > c   \\
  b \lb{ R\lp{B, C} - R\lp{B, D} } < c
\end{cases},
\end{equation}
These results perfectly align with the previous characterization of stable second-order norms in Ref.~\cite{murase2023indirect}.

\subsection*{Simultaneous observation model}

Next, we consider a model where multiple observers may assess the donor simultaneously.
To this end, we build on the work of Fujimoto and Ohtsuki~\cite{fujimoto2022reputation,fujimoto2023evolutionary,fujimoto2024leader}.
They approximate the distribution of the goodness in the stationary state for $q \!=\! 1$.
For example, when errors are rare and the population is large,  $\mu_a \!\ll\! 1$ and $N \!\to\! \infty$, the distribution of goodness under the L3 norm can be written as the sum of delta functions (See Fig.~3D in Ref.~\cite{fujimoto2023evolutionary}):
\begin{equation}
  \begin{split}
  p(g) &= (1-2\mu_a) \delta\lp{ g - \lp{1 - \mu_a} } \\
       &\quad + \mu_a \delta\lp{ g - \lp{1 - 3 \mu_a} } + \mu_a \delta\lp{g - 2\mu_a} + O(\mu_a^2).
  \end{split}
  \label{eq:goddness_dist_L3_simultaneous}
\end{equation}
From Eq.~(\ref{eq:goddness_dist_L3_simultaneous}), $h$ and $h_G$ are derived as
\begin{equation}
  h = 1 - 2\mu_a + O(\mu_a^2), \quad h_G = 1 - \mu_a + O(\mu_a^2).
  \label{eq:h_hG_simultaneous}
\end{equation}
It then follows from Eqs.~(\ref{eq:mut_best_action}) and (\ref{eq:h_hG_simultaneous}) that conditional cooperation is stable if and only if both of the following conditions are met,
\begin{equation}
\begin{cases}
  \alpha_G > 0 & \iff \frac{b}{c} > 1 + O(\mu_a)\\
  \alpha_B < 0 & \iff \frac{b}{c} < 2 + O(\mu_a).
\end{cases}
\end{equation}
Again, these conditions reproduce the results by Fujimoto and Ohtsuki~\cite{fujimoto2023evolutionary,fujimoto2024leader}.
In addition, we consider other general cases in the {\SI}.
In particular, we consider strategies that interpolate between L3 and L6, and results for $q \!<\! 1$.
For these cases, we rely on numerical simulations to calculate $h$ and $h_G$.
Once $h$ and $h_G$ are obtained, the stable range of $b/c$ is derived using Eq.~(\ref{eq:mut_best_action}).
The resulting theoretical predictions again agree with the results of numerical simulations.

\subsection*{Gossiping model}

Another application of the above theory is the gossiping model by Kawakatsu et al.~\cite{kawakatsu2024mechanistic}.
In their model, the gossip duration $\tau$ quantifies the amount of peer-to-peer gossip between private observation periods.
Kawakatsu et al. derive an analytic relationship of $h$, $h_G$ and $\tau$ as follows:
\begin{equation}
  1 - h_G = (1 - h) e^{-\tau}.
  \label{eq:h_hG_relation_gossip}
\end{equation}
When $\tau \!\to\! 0$, the gossiping process is equivalent to the solitary observation model. 
When $\tau \!\to\! \infty$, it reproduces the public assessment model.

For an illustration, we consider the assessment rule L6 (for more details, see \SI). 
Kawakatsu et al~\cite{kawakatsu2024mechanistic} consider the replicator dynamics when L6 competes with both ALLC and ALLD.
They show that a pure L6 population can only be stable when $b/c > \lp{b/c}^{\ast} \equiv 1/\lp{1-2\mu_a}$.
If that condition is satisfied, they obtain the critical gossip duration $\tau^{\ast}$ above which a pure L6 population is stable,
\begin{equation}
  \tau^{\ast} = \log \lb{ \lp{ 2 - \frac{ \boverc }{ \boverc - \frac{1}{2\lp{1 - \mu_a} }} } \lp{ \frac{ \boverc }{ \boverc - \frac{1}{1-2\mu_a} } } }.
  \label{eq:critical_gossip_duration}
\end{equation}
The same conclusion can be derived from our framework.
From Eq.~(\ref{eq:h_ast_correlated}) and Eq.~(\ref{eq:h_hG_relation_gossip}), $h$ and $h_G$ are uniquely determined.
The conditionally cooperative action rule is stable when $\alpha_G \!>\! 0$ and $\alpha_B \!<\! 0$, as defined in Eq.~(\ref{eq:alpha_G_B_definition}).
By analyzing these conditions we obtain the same critical benefit-to-cost ratio, and the same critical gossip duration.
While the previous study considers only ALLC and ALLD as possible evolutionary competitors, we conclude that the conditionally cooperative strategy is stable against mutants with any action rule, including stochastic ones.

\section*{Discussion}

In this paper, we propose a general framework to analyze the evolutionary stability of indirect reciprocity.
The literature on indirect reciprocity is vast, and researchers have established several distinct types of models~\cite{ohtsuki2004should,ohtsuki2006leading,pacheco2006stern,santos2016social,santos2018social,ohtsuki2007global,masuda2012ingroup,berger2011learning,nakamura2014indirect,ohtsuki2015reputation,clark2020indirect,murase2022social,murase2023indirect,ohtsuki2004should,ohtsuki2006leading,murase2023indirect,uchida2010effect,uchida2013effect,perret2021evolution,hilbe2018indirect,schmid2021evolution,schmid2021unified,schmid2023quantitative,fujimoto2022reputation,fujimoto2023evolutionary,fujimoto2024leader,lee2021local,okada2017tolerant,okada2018solution,okada2020two,kawakatsu2024mechanistic,kessinger2023evolution,radzvilavicius2019evolution,radzvilavicius2021adherence,krellner2021pleasing}.
Unfortunately, the different model types are often studied in isolation, and they sometimes lead to conflicting results. 
Here, we show that all this previous work can be organized by considering a single key quantity: the degree to which individual opinions are correlated.
This correlation in turn  depends on the social norm in place, on the observability of interactions, and on the degree to which individuals share their views.
As a rule of thumb we find that the more opinions are correlated, the easier it becomes to sustain cooperation. 
Conversely, if opinions turn out to be completely uncorrelated, cooperative norms become evolutionarily unstable. 
Some previous work has already hinted at the negative effects of disagreements on cooperation~\citep[e.g.][]{uchida2010effect,uchida2013effect,perret2021evolution,hilbe2018indirect}. 
Yet there has been little work to quantify these effects. 
Our study highlights the role of opinion synchronization in a mathematically explicit manner.

Within our framework, one extreme case is the public assessment model, where opinions are perfectly synchronized.
The other extreme is the solitary observation model, where opinions are statistically independent.
Although these two extreme cases may be strong idealizations, they serve as useful benchmarks due to their analytical tractability.
Between these extremes, there are several models in which opinions are correlated, but incompletely so. 
These incomplete correlations can arise, for example, when several individuals tend to witness the same event, as in the simultaneous observation model~\cite{hilbe2018indirect,schmid2021evolution,schmid2021unified,schmid2023quantitative,fujimoto2022reputation,fujimoto2023evolutionary,fujimoto2024leader,lee2021local}. 
Alternatively, they can arise when individuals use gossip to partly synchronize their views, or at least the information they have~\cite{kawakatsu2024mechanistic}. 
These intermediate models are more realistic, but they render analytical solutions more difficult to obtain.
Our results indicate that we do not need to understand the respective image matrices in full detail.
Instead we only need to know the two quantities $h$ and $h_G$ (the first two moments of the goodness distribution). 
They provide all the information needed to characterize whether a given social norm can sustain cooperation.
Future theoretical studies on opinion synchronization, such as those in Refs.~\cite{krellner2023we,kawakatsu2024mechanistic}, can provide a more detailed understanding of how the values of $h$ and $h_G$ depend on the exact social setup in which interactions take place. 

Herein, we show that some degree of opinion synchronization is crucial to maintain cooperative relationships. 
This may have implications for our understanding of typical group sizes in human societies, such as those indicated by Dunbar's number~\cite{dunbar1992neocortex,dunbar2009social}.
In smaller populations, opinions tend to synchronize more readily, thereby facilitating cooperation.
In contrast, as a group increases, opinion synchronization becomes progressively more challenging.
This difficulty in aligning opinions in larger groups ultimately sets a limit on the group size within which cooperation can be maintained.
Our research implies the importance of group size in stabilizing cooperative interactions via indirect reciprocity.

Our findings on the importance of opinion synchronization also resonate with previous work on effective punishment after norm violations~\citep{Dalkiran:SAGT:2012,Hoffman:Book:2022}. 
This literature studies under which conditions group members are able to sanction certain offenses in the presence of uncertainty. 
The model of Dalkiran et al~\citep{Dalkiran:SAGT:2012} suggests that they can only do so if it becomes common knowledge (or more precisely, `common $p$-belief') that a norm violation occurred in the first place. 
Their result follows a similar logic as in our model. 
Once individuals disagree on whether or not a norm violation occurred, it becomes too costly for any single individual to take action. 

Without a doubt, our study has several limitations.
While we have shown that opinion synchronization is crucial for evolutionarily stable indirect reciprocity, this does not mean that opinion synchronization is sufficient.
In particular, we have only considered mutants who deviate from a population's norm by choosing different actions. 
Instead, mutants may also differ in how they assess other people's actions. 
For instance, a mutant with an assessment rule that leads to a higher level of synchronization might be able to invade.
Furthermore, while opinion synchronization does help stabilize cooperation, the lack of evolutionarily stable strategies in the solitary observation model does not rule out cooperation entirely. 
Even if opinions are uncorrelated, indirect reciprocity might evolve if additional mechanisms for cooperation are in place, such as group structure~\cite{murase2024computational}.
Another valuable direction is to explore models with continuous degrees of cooperation~\citep{lee2021local,lee2022second} and models with an explicit punishment option~\citep{ohtsuki2009indirect,Schlaepfer:PRSB:2018,murase:arxiv:2024}. 

Finally, an important open question is how mechanisms for opinion synchronization co-evolve with social norms.
We studied models with simultaneous observation and gossiping, but other possibilities exist.
These mechanisms could have coexisted or evolved in a specific sequence.
Furthermore, these actions to promote synchronization may themselves be costly, which could again affect the stability of cooperation.
Investigating the evolution of these mechanisms is another promising direction for future research on indirect reciprocity.

\section*{Materials and Methods}
\subsection*{Numerical simulations}

We conducted Monte Carlo simulations to validate the theoretical predictions.
We consider a population of $N$ players. The reputation state is represented by an image matrix of size $N \times N$.
At each time step, a randomly chosen player $i$ is selected as the donor, and a randomly chosen player $j$ is selected as the recipient.
The donor $i$ decides its action based on its action rule.
Then, the reputation of the donor ($i$'th column of the image matrix) is updated.
How the reputation is updated depends on the assessment rule and the model.
More details and pseudo codes for each of these models are described in {\SI}.

We first conducted $t_{\rm init}$ steps to equilibrate the image matrix and then ran $t_{\rm measure}$ steps to measure the quantities.
The values of $h$ and $h_G$ are calculated by measuring the average and the variance of the goodness of the players~\cite{fujimoto2022reputation,fujimoto2023evolutionary}.
Here, we define the goodness of player-$i$, $g_i$, as the fraction of the good image in the $i$-th column of the image matrix excluding the diagonal element:
\begin{equation}
g_i = \sum_{j \neq i} \delta(m_{ji}, G) / (N-1),
\end{equation}
where $\delta(x, y)$ is the Kronecker delta function.
The average goodness taken over $i$ equals $h$: $h = \langle g_i \rangle$.
The product $h h_G$ is the expected probability that two randomly chosen players agree that another randomly chosen player is $G$.
Thus, for a finite $N$,
\begin{equation}
\begin{split}
h h_G &= \left\langle g_i \frac{\lp{N-1}g_i - 1}{N-2} \right\rangle = \frac{N-1}{N-2} \left\langle g_i^2 \right\rangle - \frac{\left\langle g_i \right\rangle}{N-2}  \\
  h_G &= \frac{1}{N-2} \lb{ \lp{N-1} \frac{\left\langle g_i^2 \right\rangle}{\left\langle g_i \right\rangle} - 1 }.
\end{split}
\label{eq:h_hG_from_g}
\end{equation}
When $N \to \infty$, $h_G \to \langle g_i^2 \rangle / \langle g_i \rangle$.

The parameters used in this study are $N = 100$, $\mu_a = 0.02$, $t_{\rm init} = 10^5$, and $t_{\rm measure} = 10^6$.
The source code used in this study is available at \url{https://github.com/yohm/sim_indirect_opinion_sync}.

\section*{Acknowledgments}
  Y.M. appreciates Max Planck Institute for Evolutionary Biology for its hospitality during the completion of this work.
  Y.M. acknowledges support by JSPS KAKENHI Grant Numbers JP21K03362, JP21KK0247, JP23K22087.
  C.H. acknowledges generous funding from the European Research Council (ERC)
under the European Union's Horizon 2020 research and innovation program (Starting Grant 850529: E-DIRECT), and from the Max Planck Society.

\bibliographystyle{unsrtnat}
\bibliography{indirect}

\newpage
% Supplementary Information Title Page
\begin{center}
  \LARGE Supplementary Information\\[1cm]
  \large Indirect Reciprocity under Opinion Synchronization\\[0.5cm]
  \normalsize Yohsuke Murase, Christian Hilbe\\[0.5cm]
  \today
\end{center}

\newpage  % Start SI content on a new page

% Reset section numbering for SI
\setcounter{section}{0}
\renewcommand{\thesection}{S\arabic{section}}  % Label sections as S1, S2, ...

\newpage

\section{Models}

\subsection{Models of indirect reciprocity}

We consider a population of size $N (\gg 1)$ in which each player is engaged in the donation game repeatedly.
In each round, a player is randomly selected as a donor, and another one is selected randomly as a recipient.
The donor chooses either to cooperate ($C$) or defect ($D$).
By cooperating, the donor pays a cost $c$ to benefit the recipient by $b (> c)$. By defecting, players get no benefit or pay no cost.
The game is repeated for sufficiently many rounds changing the donor and recipient in each round, and the players accumulate their payoffs over the rounds.

During this sequence of the donation games, players form opinions about each other.
The opinion player $i$ holds about $j$ is denoted as $m_{ij}$, and the entire matrix $M$ is referred to as an image matrix.
Following the convention of previous literature, opinions are binary, either good ($G$) or bad ($B$), and they can change over time.
Donor players condition their actions on the opinions they hold about the recipient.
The way they decide whether to cooperate in the donation game is determined by their action rule $P$.
The action is observed by other players, who then update their opinions about the donor with their assessment rules $R$.
The combination of these rules $\{P, R\}$ is referred to as the strategy or the social norm.

We consider the so-called stochastic `second-order' norms.
Donor $i$ decides its action $A_{ij} (\in \{C, D\})$ based on their opinions about the recipient $j$. The donor cooperates with the probability prescribed by the action rule $P(m_{ij})$.
The assessment by an observer $k$ on the donor $i$ is determined by the action taken by the donor and the observer's opinion about the recipient.
The assessment rule $R$ is a function of $m_{kj}$ and $A_{ij}$ that returns the probability that the donor is assessed as $G$.
The new opinion about the donor $m_{ki}'$ is $G$ with probability $R(m_{kj}, A_{ij})$ otherwise $m_{ki}' = B$.
Since neither $P$ nor $R$ depends on the self-image, $m_{ii}$ does not play any role in this model.

Both the players' actions and their assessments can be subject to errors.
First, we consider implementation errors. Such errors affect donors who wish to cooperate; with probability $\mu_e$ such donors defect by mistake.
As a result, instead of their intended strategy $P(X)$, players implement the effective strategy
\begin{equation}
  \tilde{P}(X) = \lp{1-\mu_e} P(X).
\end{equation}
In this paper, we consider $\mu_e = 0$ unless explicitly stated because the dynamics are not qualitatively different from the case of $\mu_e = 0^{+}$.
Similarly, we consider assessment errors. These errors happen when new reputations are assigned. With probability $\mu_a$, the assignment is the opposite of those prescribed by the social norm.
As a result, the effective assessment rules become
\begin{equation}
  \tilde{R}\lp{X, A} = \lp{1-\mu_a}R\lp{X, A} + \mu_a \lb{1-R\lp{X, A} }.
\end{equation}
In the following, we denote $\tilde{R}$ as $R$ and $\tilde{P}$ as $P$ for simplicity unless explicitly stated.
In this paper, we limit ourselves to the cases that $\mu_a > 0$, i.e., $0 < R(X, A) < 1$ for any $X$ and $A$.
This condition is crucial to make the dynamics ergodic. Irrespective of the initial conditions, the system reaches a unique stationary state after a long enough time.

We consider four models of indirect reciprocity: the solitary observation model, the public assessment model, the simultaneous observation model, and the gossiping model.
They differ in how the opinions are updated and how the assessments are made.
The common part of the models, except for the gossiping model, is described by the pseudo-code in Algorithm~\ref{alg:main}.
In each time step, a donor $i$ and a recipient $j$ are randomly chosen.
The donor decides its action $C$ or $D$ based on its action rule $P_i(m_{ij})$.
Then, the opinions are updated by a function $\text{UpdateOpinions}(i, j)$, which is model-dependent, as shown in the following subsections.

\begin{algorithm}[h]
  \caption{
    A pseudo-code of a time step for the indirect reciprocity models. The function $\text{UpdateOpinions}(i, j)$ updates the opinions based on the action taken by $i$ toward $j$.
    Implementation of the function $\text{UpdateOpinions}(i, j)$ is model-dependent.
  }
  \label{alg:main}
  \begin{algorithmic}[1]
  \Procedure{Main}{}
    \For{$t_{init}$ times}
      \State \Call{ConductGame}{}  \Comment{Initial thermalization}
    \EndFor
    %\State Reset the counters
    \For{$t_{max}$ times}
      \State \Call{ConductGame}{}   \Comment{Main simulation}
    \EndFor
    \State Calculate the cooperation probabilities between players
  \EndProcedure

  \Procedure{ConductGame}{}
    \State $i \gets \text{Randomly select a donor}$
    \State $j \gets \text{Randomly select a recipient other than } i$
    \State $coopProb \gets P_i(m[i][j])$  \Comment{Get the cooperation probability using $i$'s action rule}
    \State $r \gets \text{Random}(0, 1)$
    \If{$r < coopProb$}
      \State $action \gets C$   \Comment{Cooperation}
      % \State $costCount[i] \gets costCount[i] + 1$
      % \State $benefitCount[j] \gets benefitCount[j] + 1$
    \Else
      \State $action \gets D$   \Comment{Defection}
    \EndIf
    % \State $donorCount[i] \gets donorCount[i] + 1$
    % \State $recipCount[j] \gets recipCount[j] + 1$
    \State \Call{UpdateOpinions}{$i$, $j$, $action$}
  \EndProcedure
\end{algorithmic}
\end{algorithm}

\begin{algorithm}[h]
  \caption{
    A pseudo-code of $\text{UpdateOpinions}(i, j)$ for the solitary observation model.
  }
  \label{alg:solitary}
  \begin{algorithmic}[1]
  \Procedure{UpdateOpinions}{$i$, $j$, $action$}
    \State $k \gets \text{Randomly select an observer other than } i, j$
    \State $goodProb \gets R_k(m[k][j], action)$  \Comment{Assess using $k$'s assessment rule}
    \State $r \gets \text{Random}(0, 1)$
    \If{$r < goodProb$}
      \State $m[k][i] \gets G$
    \Else
      \State $m[k][i] \gets B$
    \EndIf
  \EndProcedure
\end{algorithmic}
\end{algorithm}

\begin{algorithm}[h]
  \caption{
    A pseudo-code of $\text{UpdateOpinions}(i, j)$ for the public assessment model.
    Note that the opinions are always fully synchronized by model assumption. Namely, each column of the image matrix is always identical.
    All players use the same assessment rule.
  }
  \label{alg:public}
  \begin{algorithmic}[1]
  \Procedure{UpdateOpinions}{$i$, $j$, $action$}
    \State $goodProb \gets R(m[i][j], action)$
    \State $r \gets \text{Random}(0, 1)$
    \If{$r < goodProb$}
      \For{$k \gets 0 \text{ to } N-1$}
        \State $m[k][i] \gets G$
      \EndFor
    \Else
      \For{$k \gets 0 \text{ to } N-1$}
        \State $m[k][i] \gets B$
      \EndFor
    \EndIf
  \EndProcedure
\end{algorithmic}
\end{algorithm}

\begin{algorithm}[h]
  \caption{
    A pseudo-code of $\text{UpdateOpinions}(i, j)$ for the simultaneous observation model.
    All players update their opinions about $i$ simultaneously using their own assessment rule with observation probability $q$.
  }
  \label{alg:simultaneous}
  \begin{algorithmic}[1]
  \Procedure{UpdateOpinions}{$i$, $j$, $action$}
    \For{$k \gets 0 \text{ to } N-1$}
      \State $r \gets \text{Random}(0, 1)$
      \If{$r < q$}
        \State $goodProb \gets R_k(m[k][j], action)$
        \State $r2 \gets \text{Random}(0, 1)$
        \If{$r2 < goodProb$}
          \State $m[k][i] \gets G$
        \Else
          \State $m[k][i] \gets B$
        \EndIf
      \EndIf
    \EndFor
  \EndProcedure
\end{algorithmic}
\end{algorithm}

\begin{algorithm}[htbp]
  \caption{
    A pseudo-code for the gossiping model.
    We use a different $\text{ConductGame}$ function from the other models.
  }
  \label{alg:gossiping}
  \begin{algorithmic}[1]
  \Procedure{ConductGame}{}
    \For{$i \gets 0 \text{ to } N-1$}  \Comment{Interaction phase}
      \For{$j \gets 0 \text{ to } N-1$}
        \If {$i = j$}
          \State \textbf{continue}
        \EndIf
        \State $coopProb \gets P_i(m[i][j])$
        \State $r \gets \text{Random}(0, 1)$
        \If{$r < coopProb$}
          \State $actions[i][j] \gets C$
        \Else
          \State $actions[i][j] \gets D$
        \EndIf
      \EndFor
    \EndFor

    \For{$i \gets 0 \text{ to } N-1$}  \Comment{Assessment phase}
      \For{$j \gets 0 \text{ to } N-1$}
        \If {$i = j$}
          \State \textbf{continue}
        \EndIf
        \State $k \gets \text{Randomly select a player other than } i$
        \State $goodProb \gets R_i(m[i][k], actions[j][k])$
        \State $r \gets \text{Random}(0, 1)$
        \If{$r < goodProb$}
          \State $assessments[i][j] \gets G$
        \Else
          \State $assessments[i][j] \gets B$
        \EndIf
      \EndFor
    \EndFor

    \For{$i \gets 0 \text{ to } N-1$}   \Comment{New assessments are assigned to the image matrix}
      \For{$j \gets 0 \text{ to } N-1$}
        \If {$i = j$}
          \State \textbf{continue}
        \EndIf
        \State $m[j][i] \gets assessments[j][i]$
      \EndFor
    \EndFor

    \For {$n \gets 0 \text{ to } n_{\text{gossip}}-1$}  \Comment{Gossiping phase}
      \State $i \gets \text{Randomly select a player}$
      \State $j \gets \text{Randomly select a player other than } i$
      \State $k \gets \text{Randomly select a player other than } i, j$
      \State $m[i][k] \gets m[j][k]$
    \EndFor
  \EndProcedure

\end{algorithmic}
\end{algorithm}

\subsection{Solitary observation model}

The first model we consider is the solitary observation model~\cite{okada2018solution}.
In this model, a randomly selected observer $k$ updates the opinion about the donor $i$ based on the action taken by the donor $i$ toward the recipient $j$.
The observer $k$ assesses the donor $i$ using its assessment rule $R_k$ and its opinion about the recipient $m_{kj}$.
The pseudo-code is shown in Algorithm~\ref{alg:solitary}.

Alternatively, we may consider an equivalent model.
Each time step consists of an interaction phase and an opinion update phase.
In the interaction phase, for every pair of players $i$ and $j$, the donor $i$ conducts its action based on its action rule $P_i(m_{ij})$.
After the interaction phase, the opinions are updated based on the actions taken by the donors.
A player $k$ updates the opinion about $i$, $m_{ki}$, based on a randomly selected action that $i$ took.
Let's say the action was conducted toward $j$.
The opinion is updated based on the assessment rule $R_k(m_{kj}, A_{ij})$.
In an opinion update phase, all elements of the image matrix are updated.

These two interpretations are essentially equivalent.
The point is that $O(N^2)$ games are conducted to update $O(N^2)$ elements of the entire image matrix.
Each action is observed only once by a single observer on average.
This way, elements in each column are statistically independent, and thus analytical calculations are straightforward.

\subsection{Public assessment model}

In the public assessment model, everyone fully agrees on their opinions.
Namely, each column of the image matrix is always identical.
This is the opposite extreme of the solitary observation model.
Since the opinions are always synchronized, the population has a single assessment rule $R$.
A pseudo-code of the opinion update is shown in Algorithm~\ref{alg:public}.

This assumption simplifies the analysis of the reputation dynamics.
In the literature, this model has been most intensively studied after a seminal work by Ohtsuki and Iwasa~\cite{ohtsuki2004should} on the leading eight.
In one of those studies, Murase and Hilbe~\cite{murase2023indirect} generalized the model of Ohtsuki and Iwasa to incorporate stochastic action and assessment rules.
For the generalized model, they identified the cooperative ESSs comprehensively by deriving the necessary and sufficient conditions for the social norms to be evolutionarily stable.
In this paper, we will use this stochastic version of the model to compare the results with the other models.

\subsection{Simultaneous observation model}

In the simultaneous observation model, a single action is observed by other players, and the opinions about the donor are updated simultaneously~\cite{hilbe2018indirect}.
Unlike the solitary observation model, the number of observers is not limited to one.
All players except the donor themselves observe the action $A_{ij}$ with observation probability $q$.
Each observer makes the assessments independently.
Let us say $k$ is the observer.
This observer $k$ updates the opinion about the donor $m_{ki}$ based on the action $A_{ij}$ and the observer's opinion about the recipient $m_{kj}$.
With probability $R_k(m_{kj}, A_{ij})$, $m_{ki}$ is updated to $G$. Otherwise, it is updated to $B$.
Since the opinions toward the recipient may differ between observers and each assessment is subject to errors, the new opinions are not necessarily shared between the observers, even if they use the same assessment rule.
This model includes the solitary observation model as a special case when $q = 1/N$.
A pseudo-code of the opinion update is shown in Algorithm~\ref{alg:simultaneous}.

While this model is more realistic because multiple observers can assess the same action, it is more challenging to analyze than the solitary observation model since the opinions are not statistically independent.
In general, it is hard to obtain analytical solutions for the dynamics of the opinions except for a few special cases~\cite{fujimoto2022reputation,fujimoto2023evolutionary}.
Simulations are required to investigate the dynamics of the opinions and the cooperation probabilities.

\subsection{Gossiping model}

Lastly, we consider a model of gossiping proposed by Kawakatsu et al.~\cite{kawakatsu2024mechanistic}.
This model uses an alternative version of the solitary observation model as a base.
Namely, at each time step, donation games are played by all pairs of players.
After the interaction phase, the opinions are updated.
Each assessment is conducted according to an action randomly selected from the donor's actions.
In addition to these interaction and the assessment phases, the gossiping phase is introduced, where players exchange their opinions with randomly selected peers.
In a gossipnig event, three players $i$, $j$, and $k$ are randomly selected, and $m_{ij}$ is copied to $m_{kj}$.
This event is repeated $\tau N^3 / 2$ times in a gossiping phase, where $\tau$ is the gossiping duration.
A pseudo-code is shown in Algorithm~\ref{alg:gossiping}.
A source code is provided by the authors of the original paper~\cite{kawakatsu2024mechanistic}.

This model falls back to the solitary observation model when $\tau = 0$.
Conversely, when $\tau \to \infty$, the opinions are fully synchronized, thus equivalent to the public assessment model.

\newpage
\section{Analysis on the solitary observation model}

\subsection{Conditional cooperation does not form ESS}

In the following, we are going to demonstrate that the conditional cooperator does not generally form ESS in the solitary observation model.
This is a more detailed version of the argument in the main text.
An intuition is the following:
We consider the best action for the donor (Alice, $\mathcal{A}$) against the recipient (Bob, $\mathcal{B}$) under the observation by Charlie ($\mathcal{C}$).
In the case of the solitary observation, the opinions toward Bob from Alice and from Charlie are statistically independent.
In order to maximize the payoff, Alice does not have to change her actions toward Bob according to her opinion because only Charlie's opinion matters for Alice's long-term payoff.
If the long-term benefit of cooperation exceeds the immediate cost of cooperation, Alice should always cooperate irrespective of her opinion.
Conversely, Alice should defect unconditionally when the cost of cooperation is higher.
Namely, Alice does the free-riding or the second-order free-riding to maximize her payoff.

To support this intuition, let us consider the best response more formally.
Assume a sufficiently large population, in which each action is observed by a single observer only.
In this case, the probability that $j$ is assessed as $G$ is independent of the observers, $m_{ij}$ is statistically independent of $m_{kj}$ where $i \neq k$.
Namely, the image $m_{ij} = 1$ with probability $h$ otherwise $0$ for any $i$, where $h$ is the expected fraction of $G$ image in the image matrix.

First, we consider a mono-morphic resident population in which everyone uses the same norm $\{P, R\}$.
Let us denote the average fraction of $G$ in the image as $h$.
Because of the statistical independence, any element of the image matrix $m_{ij}$ is $G$ with probability $h$ and $B$ otherwise.
The cooperation probability of the population is written as
\begin{equation}
p_c = hP(G) + (1-h)P(B).
\label{eq:p_c}
\end{equation}
The probability that Charlie assesses Alice as $G$ after his observation is
\begin{equation}
h \lb{ p_c R\lp{G,C} + \lp{1-p_c} R\lp{G,D} } + \lp{1-h} \lb{ p_c R\lp{B,C} + \lp{1-p_c} R\lp{B,D} }.
\end{equation}
Therefore, the dynamics of $h$ is described as follows:
\begin{equation}
  \dot{h} = h^2 R_P\lp{G, G} + h(1-h) \lb{ R_P\lp{G, B} + R_P\lp{B, G} } + \lp{1-h}^2 R_P\lp{B, B} - h,
  \label{eq:h_dot_solitary}
\end{equation}
where we define
\begin{equation}
R_P(X, Y) \equiv P(X)R(Y,C) + \lp{ 1-P\lp{X} } R(Y,D).
\label{eq:R_P}
\end{equation}
This is the probability that the donor Alice is assessed as $G$ by the observer Charlie given their original opinions toward the recipient Bob are $m_{\mathcal{AB}} = X$ and $m_{\mathcal{CB}} = Y$, respectively.
The first term of \eqref{eq:h_dot_solitary} corresponds to the case when both Alice and Charlie assess Bob as $G$ before the interaction.
The other terms correspond to the cases for $(m_{\mathcal{AB}}, m_{\mathcal{CB}}) = (G, B)$, $(B, G)$, and $(B, B)$.

When the assessment error rate $\mu_a$ is not zero, $h$ reaches a unique stationary solution $h^{\ast}$.
The unique stationary solution $h^{\ast}$ is obtained by $\dot{h} = 0$:
\begin{equation}
  h^{\ast} = \begin{cases}
    \frac{ -c_1 - \sqrt{c_1^2 - 4c_2c_0}}{ 2c_2 } & (c_2 \neq 0)  \\
    - \frac{c_0}{c_1}  & (c_2 = 0)
    \end{cases},
\end{equation}
where
\begin{equation}
\begin{split}
  c_2 &\equiv R_P(G,G) - R_P(G,B) -  R_P(B,G) + R_P(B,B)   \\
  c_1 &\equiv R_P(G,B) + R_P(B,G) - 2R_P(B,B) - 1   \\
  c_0 &\equiv R_P(B,B)
\end{split}.
\end{equation}
Therefore, when a second-order social norm is given, the average reputation $h^{\ast}$ and the self-cooperation level are obtained as above.
The condition for $h^{\ast} = 1$ is obtained as
\begin{equation}
  h^{\ast} = 1  \iff \begin{cases}
    R_P(G,G) = 1  \\
    R_P(G,B) + R_P(B,G) > 1
  \end{cases}
\end{equation}
In the following, we consider only the stationary state so we denote $h^{\ast}$ as $h$ for simplicity.

Next, let us consider the best response.
Assume that a small amount of a mutant species having a different norm comes into the system.
Suppose that the average cooperation rate of the mutant toward the residents is $p_{\rm mut \to res}$.
Here, we do not have to define the specific form of the action and the assessment rules for the mutant species because only the mutant's actions matter for the residents to assess the mutant.
The mutant may have a complex norm but the mutant is inaccessible to the resident's opinion anyway.
Similarly, the residents do not know what the mutant considers. They exchange information only by their actions.

The expected image about the mutant player Alice from the residents $H$ follows the dynamics as
\begin{equation}
  \dot{H} = p_{\rm mut \to res} \lb{ h R(G,C) + \lp{1-h} R(B,C) } + \lp{1 - p_{\rm mut \to res}} \lb{ hR(G,D) + \lp{1-h} R(B,D) } - H.
\end{equation}
After a sufficiently long time, $H$ converges to its stationary value
\begin{equation}
  H^{\ast} = p_{\rm mut \to res} \lb{ h R(G,C) + \lp{1-h} R(B,C) } + \lp{1 - p_{\rm mut \to res}} \lb{ hR(G,D) + \lp{1-h} R(B,D) }.
\end{equation}
Below, we denote $H^{\ast}$ as $H$ for simplicity.
The cooperation probability from the resident to the mutant is
\begin{equation}
\begin{split}
  p_{\rm res \to mut} &= H P(G) + (1-H) P(B)   \\
    &= \lc{ p_{\rm mut \to res} \lb{ h R(G,C) + \lp{1-h} R(B,C) } + \lp{1 - p_{\rm mut \to res}} \lb{ hR(G,D) + \lp{1-h} R(B,D) } } P(G) \\
    &\quad + \lc{ p_{\rm mut \to res} \lb{ h \bar{R}(G,C) + \lp{1-h} \bar{R}(B,C) } + \lp{1 - p_{\rm mut \to res}} \lb{ h\bar{R}(G,D) + \lp{1-h} \bar{R}(B,D) } } P(B) \\
    &= p_{\rm mut \to res} P_b + P_0,
\end{split}
\label{eq:pc_res_mut}
\end{equation}
where $\bar{R}(X,Y) = 1 - R(X,Y)$.
Therefore, $p_{\rm res \to mut}$ is a linear function of $p_{\rm mut \to res}$. Its slope and intercept are
\begin{equation}
  P_{b} \equiv \lb{P\lp{G} - P\lp{B} } \lc{ h \lb{R\lp{G,C} - R\lp{G,D} } + \lp{1-h} \lb{ R\lp{B,C} - R\lp{B,D} } },
  \label{eq:P_b}
\end{equation}
and
\begin{equation}
  P_0 \equiv h\lc{ R\lp{G,D} P\lp{G} + \lb{1 - R\lp{G,D} }P\lp{B} } + \lp{1-h} \lc{ R\lp{B,D} P\lp{G} + \lb{1 - R\lp{B,D} }P\lp{B} },
\end{equation}
respectively.

We verified this linear relationship in Eq.~(\ref{eq:pc_res_mut}) with numerical simulations as shown in the main text.
This linearity is not limited to the cases when the mutants are the stochastic unconditional cooperators but also holds for more complex classes of strategies, such as the second-order norms.
The payoff of the mutant $\pi_{\rm mut}$ is written as
\begin{equation}
\begin{split}
  \pi_{\rm mut} &= bp_{\rm res \to mut} - cp_{\rm mut \to res}   \\
                &= \lb{b P_b - c} p_{\rm mut \to res} + \lp{\text{terms independent of $p_{\rm mut \to res}$}}
                %&+ bh^{\ast}\{ R(G,D)P(G) + [1 - R(G,D)]P(B) \} + (1-h^{\ast}) \{ R(B,D)P(G) + [1 - R(B,D)]P(B) \},
\end{split}
\end{equation}
The mutant maximizes its payoff when
\begin{equation}
  p_{\rm mut \to res} = \begin{cases}
    1 & {\rm when \quad} b P_{b} > c  \\
    {\rm any} & {\rm when \quad} b P_{b} = c \\
    0 & {\rm when \quad} b P_{b} < c
  \end{cases}.
\end{equation}
Thus, a mutant can maximize its payoff by unconditionally cooperating or defecting. Conditional cooperation does not pay off in general.
In other words, the resident species cannot exclude the free riders and the second-order free riders, simultaneously.

Here, $bP_{b}$ is interpreted as the expected net benefit of cooperation.
The first factor of Eq.~(\ref{eq:P_b}), $\lb{ P\lp{G} - P\lp{B} }$, indicates the incentives of being $G$ in the residents' viewpoint.
The larger this factor, the more likely the good mutant would be cooperated compared to the case the mutant were $B$.
The second factor $\lc{ h\lb{ R\lp{G,C} - R\lp{G,D} } + \lp{1-h} \lb{ R\lp{B,C} - R\lp{B,D} } }$ indicates how more likely the mutant gets $G$ reputation by cooperating.

While the discriminator does not form a Nash equilibrium in general, $b P_b = c$ is the only exception.
In this case, the mutant can obtain the same payoff as the residents, thus forming a Nash equilibrium.
Generous Scoring (GSCO) norm proposed by Schmid et al.~\cite{schmid2021unified} corresponds to this case.
GSCO prescribes $P(G) = 1, P(B) = 0, R(-,C) = 1, R(-,D) = 1 - c/(1-2\mu_a)b$ when $N \gg 1$.
The rescaled assessment rule $\tilde{R}(-, C) = 1 - \mu_a$ and $\tilde{R}(-, D) = 1 - \mu_a - c/b$ satisfies $b P_b = c$.

The above argument shows that it is not easy to stabilize cooperation in the solitary observation model.
This essentially comes from the fact that the mutant has no information about what the observer considers about the recipient since opinions are statistically independent.
Therefore, it is essential to introduce the correlation between opinions to overcome the second-order free rider problem.

\subsection{Non-binary reputation model}

\begin{table}[h]
  \centering
  \begin{tabular}{c|cccccc|ccc}
    Norm & $R(G,C)$ & $R(G,D)$ & $R(N,C)$ & $R(N,D)$ & $R(B,C)$ & $R(B,D)$ & $P(G)$ & $P(N)$ & $P(B)$ \\
    \hline
    GBGGBN-100 & $G$ & $B$ & $G$ & $G$ & $B$ & $N$ & $1$ & $0$ & $0$ \\
    GBGGNG-100 & $G$ & $B$ & $G$ & $G$ & $N$ & $G$ & $1$ & $0$ & $0$ \\
  \end{tabular}
  \caption{
    The definition of the norms used in the ternary reputation model.
    The assessment rule $R(X, A)$ is the reputation assigned to the donor when the observer's opinion about the recipient is originally $X$ and the donor's action is $A$.
    These norms are cooperative ESSs in the public assessment model.
  }
  \label{tab:ternary_norms}
\end{table}

\begin{figure}[h]
  \centering
  \includegraphics[width=0.8\textwidth]{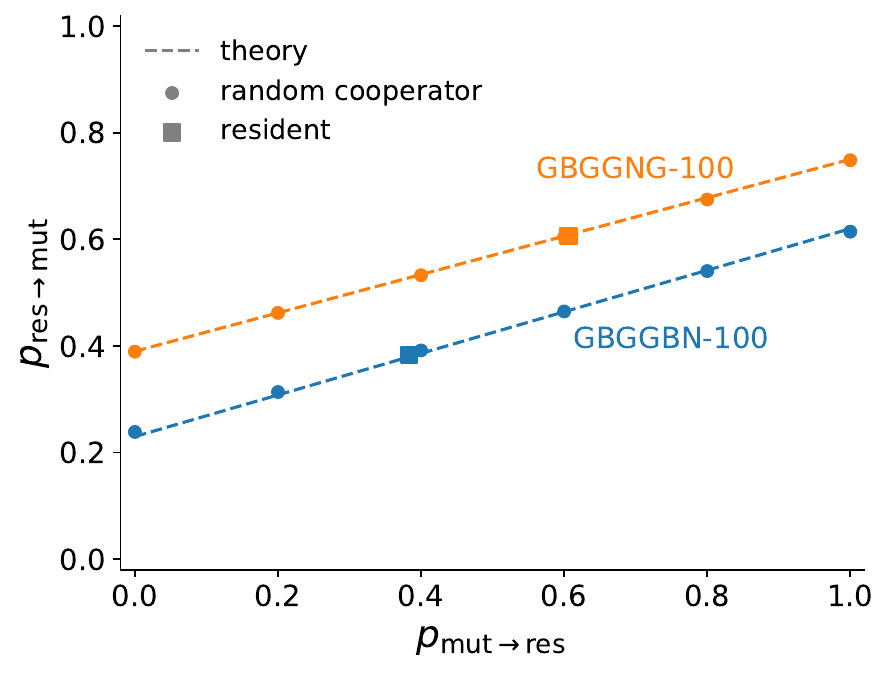}
  \caption{
    The cooperation probability of the residents toward the mutants $p_{\rm res \to mut}$ as a function of the cooperation probability of the mutants toward the residents $p_{\rm mut \to res}$ for the solitary observation model.
    The circles represent the simulation results when a single mutant, who is a stochastic unconditional cooperator, is introduced.
    The square represents the cooperation probability of the residents for a population without mutants.
    The total population size is $N=100$, and the assessment error rate is $\mu_a = 0.02$.
    The definition of the resident strategy is shown in Table~\ref{tab:ternary_norms}.
    Theoretical predictions by Eq.~(\ref{eq:pc_res_mut_general}) are shown as the dashed lines.
  }
  \label{fig:ternary_reputation}
\end{figure}

While it is a common practice to assume binary reputations, it is straightforward to generalize the conclusion that the solitary observation model cannot sustain evolutionarily stable cooperation.
Since this negative conclusion comes from the statistical independence of the opinions, it is not limited to the binary reputation model.

Consider a generalized model where reputations are represented by $M$ different states, $\lc{S_1, S_2, \ldots, S_M}$.
We define an action rule $\mathbf{P}$ as a column vector that assigns cooperation probabilities for each reputation state, $\mathbf{P} = \lp{P_1, P_2, \ldots, P_M}^\intercal$.
For instance, a donor cooperates with probability $P_i$ when its opinion about the recipient is $S_i$.
Similarly, an assessment rule is represented by a pair of $M \times M$ matrices, $\lc{R^C, R^D}$.
The $(i, j)$-th element of $R_C$, $R^C_{ij}$, is defined as the probability that the observer assesses the donor as $S_j$ after observing the donor's cooperation toward the recipient whose reputation was originally $S_i$ from the observer's viewpoint.
By definition, the row sum $\sum_j R^C_{ij} = 1$ for all $i$.
In the presence of assessment errors, all elements are neither zero nor one, $0 < R^C_{ij} < 1$.
The other matrix $R^D$ is defined similarly for defection.

The reputational state of a monomorphic population is represented by a column vector $\mathbf{h} = \lp{h_1, h_2, \ldots, h_M}^\intercal$, where $h_i$ is the fraction of $S_i$ in the off-diagonal elements of the image matrix.
Let us first consider the dynamics of its first element $h_1$. Its time evolution is
\begin{equation}
  \begin{split}
  \dot{h_1} &= \sum_{i=1}^M \sum_{j=1}^M h_i h_j \lc{ P_i R^C_{j1} + \lb{1-P_i} R^D_{j1} }  - h_1  \\
            &= \lp{\mathbf{h}\cdot\mathbf{P}} \sum_{j=1}^M h_j R^C_{j1} + \mathbf{h}\cdot\lp{\mathbf{1}-\mathbf{P}} \sum_{j=1}^M h_j R^D_{j1} - h_1,
            %&= \lp{\mathbf{h}\cdot\mathbf{P}} \lp{ R^{C\intercal} \mathbf{h} }_1 + \lb{\mathbf{h}\cdot\lp{\mathbf{1}-\mathbf{P}}} \lp{ R^{D\intercal}\mathbf{h} }_1 - h_1,
  \end{split}
\end{equation}
where $\mathbf{1}$ is the column vector of ones.
% and $(\dots)_1$ is the first element of the vector.
More generally,
\begin{equation}
  \dot{\mathbf{h}} = \lc{ \lp{\mathbf{h}\cdot\mathbf{P}} R^C + \lb{\mathbf{h}\cdot\lp{\mathbf{1}-\mathbf{P}}} R^D }^{\intercal}\mathbf{h}  - \mathbf{h}.
\end{equation}
We assume that the population reaches a stationary state $\mathbf{h}^{\ast}$ after a sufficiently long time. The stationary state satisfies
\begin{equation}
  \mathbf{h}^{\ast} = \lc{ \lp{\mathbf{h}^{\ast}\cdot\mathbf{P}} R^C + \lb{\mathbf{h}^{\ast}\cdot\lp{\mathbf{1}-\mathbf{P}}} R^D }^{\intercal} \mathbf{h}^{\ast}
\end{equation}
%\YM{I hope that the stationary state is uniquely determined. Can we prove this? We cannot use the Perron-Frobenius theorem directly. Since the matrix $\lc{\dots}$ contains $\mathbf{h}^\ast$ inside. Even if it is difficult to prove the uniqueness of the stationary state, we want to show the existence of at least one stationary state.}
Hereafter, we consider only the stationary state. We denote $\mathbf{h}^{\ast}$ as $\mathbf{h}$ for short.
The self-cooperation level is written as the inner product:
\begin{equation}
  p_{\rm res \to res} = \mathbf{h} \cdot \mathbf{P}.
\end{equation}

Next, we consider an infinitesimal amount of mutant players with a different norm.
Although a mutant may have a different (potentially arbitrarily complex) assessment and action rules, we only need to consider the expected cooperation probability to the residents.
This is because the opinion of the mutant is statistically independent of the opinions of the residents.
From the resident's viewpoint, only the mutant's cooperation probability matters.
Let us say $P'_i$ is the mutant's cooperation probability toward a recipient whose reputation is $S_i$ from a resident observer's viewpoint.
Because of the assumption in the solitary observation model, $P'_1 = P'_2 = \dots = P'_M$ since the opinions of the residents and the mutant are independent.
Let us denote this mutant's cooperation probability as $p_{\rm mut \to res}$.
The reputation of the mutant $\mathbf{H}$ is
\begin{equation}
  \mathbf{H} = \lb{ p_{\rm mut \to res} R^C + \lp{ 1 - p_{\rm mut \to res}} R^D }^{\intercal} \mathbf{h}.
\end{equation}
The cooperation probability from the resident to the mutant is
\begin{equation}
  \begin{split}
  p_{\rm res \to mut} &= \mathbf{H} \cdot \mathbf{P}  \\
    &= p_{\rm mut \to res} \lb{\mathbf{h}^{\intercal} \lp{R^C - R^D} \mathbf{P} } + \mathbf{h}^{\intercal} R^{D} \mathbf{P}
  \end{split}
  \label{eq:pc_res_mut_general}
\end{equation}
The payoff of the mutant is
\begin{equation}
  \begin{split}
  \pi_{\rm mut} &= b p_{\rm res \to mut} - c p_{\rm mut \to res}  \\
                &= p_{\rm mut \to res} \lb{ b \mathbf{h}^{\intercal} \lp{R^C - R^D} \mathbf{P} - c } + b \mathbf{h}^{\intercal} R^D \mathbf{P}.
  \end{split}
\end{equation}
Therefore, $\pi_{\rm mut}$ is a linear function of $p_{\rm mut \to res}$ thus the best action rule for the mutant is
\begin{equation}
  \begin{cases}
    p_{\rm mut \to res} = 1          & \text{if } b \mathbf{h}^{\intercal} \lp{R^C - R^D} \mathbf{P} - c > 0  \\
    p_{\rm mut \to res} = \text{any} & \text{if } b \mathbf{h}^{\intercal} \lp{R^C - R^D} \mathbf{P} - c = 0  \\
    p_{\rm mut \to res} = 0          & \text{if } b \mathbf{h}^{\intercal} \lp{R^C - R^D} \mathbf{P} - c < 0.  \\
  \end{cases}
\end{equation}
Either ALLC or ALLD is the best action rule in general and the conditional cooperation is not evolutionarily stable.

Similar to the case of the binary reputation, this is a consequence of the statistical independence of the opinions.
The donor does not have any incentive to cooperate conditionally on its own opinion about the recipient.
Because of the statistical independence, the donor's action looks like a random action from the observer's viewpoint.

The above conclusion is valid for any number of reputation states $M$.
The reputation could be represented by ternary values~\cite{murase2022social}, integers~\cite{schmid2023quantitative}, or even continuous values~\cite{lee2021local}.
Conditional cooperation cannot be evolutionarily stable if the opinions are statistically independent.

To verify the above argument, we conducted numerical simulations for the ternary reputation model~\cite{murase2022social}.
The reputations (opinions) are represented by three states, $\lc{G, N, B}$.
In the case of public assessment, cooperative ESS norms are comprehensively identified, and we use some second-order norms in Table~5 in Ref.~\cite{murase2022social}.
The definition of the norms is shown in Table~\ref{tab:ternary_norms}.
Figure~\ref{fig:ternary_reputation} shows the cooperation probability of the residents toward the mutants $p_{\rm res \to mut}$ as a function of $p_{\rm mut \to res}$.
The figure demonstrates that the relationship is linear, and the mutants can unconditionally cooperate or defect to maximize their payoff.

\subsection{Neutral invasion by a random cooperator under solitary observations}

\begin{figure}[h]
  \centering
  \includegraphics[width=0.8\textwidth]{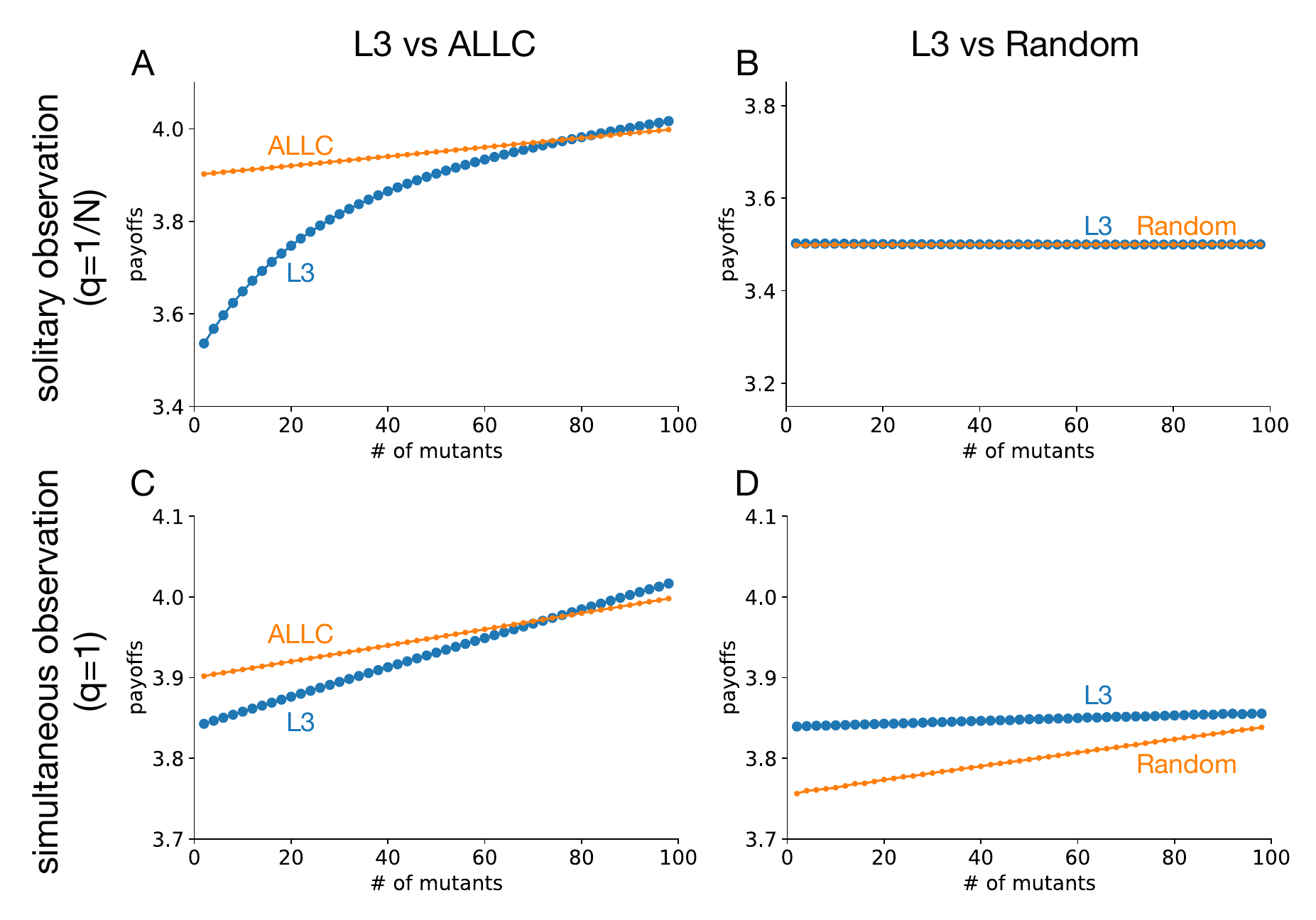}
  \caption{
    The payoff comparison between L3 residents vs (A,C) ALLC mutants and (B,D) random mutants.
    The upper panels are the simulation results for the solitary observation model while the lower panels are simulation results for the simultaneous observation model with $q = 1$.
    The total population size is $N=100$ and the assessment error rate is $\mu_a = 0.02$.
    The cooperation probabilities for the random mutants in (B,D) are the same as the average cooperation probability of the pure L3 population, which are $0.875$ and $0.96$, respectively.
  }
  \label{fig:L3_vs_mutants}
\end{figure}

The above argument is about the local stability in the vicinity of the monomorphic population.
Even when the fraction of the mutants increases, we here show that any norm cannot prevent neutral drift by a random cooperator under solitary observations.
Figure~\ref{fig:L3_vs_mutants}A shows the payoffs of L3 residents and ALLC mutants as a function of the number of mutant players.
While pure L3 is locally unstable against ALLC, L3's payoff surpasses ALLC's as the fraction of the mutants increases, yielding a stable coexistence.
By the nature of ALLC, the payoffs of ALLC mutants eventually become less than the residents' payoffs as the fraction of ALLC increases.

However, the population cannot resist neutral invasion by a random cooperator.
When the random mutants with the same cooperation probability as the residents, the residents and mutants always have the same payoff, as shown in Fig.~\ref{fig:L3_vs_mutants}B.
In other words, L3 resident players cannot distinguish other L3 players from those randomly cooperating with the same probability.
This is again because of the statistical independence of the opinions.
For an L3 observer, the action of other L3 players looks as if it were random.
Thus, the random cooperators can neutrally invade the residents.
This is not limited to L3 but is a general feature of the solitary observation model.
Thus, it is impossible to have ESS with conditional cooperation under solitary observations.

If we do the same analysis for the simultaneous observation model, the results are different.
As shown in Fig.~\ref{fig:L3_vs_mutants}C, the L3 residents have less payoff than ALLC mutants when the mutants are rare but they reach stable coexistence, similarly to the solitary observation model Fig.~\ref{fig:L3_vs_mutants}A.
However, the response to the random cooperators is different as shown in Fig.~\ref{fig:L3_vs_mutants}D.
The L3 residents have higher payoffs than the random cooperators, and the random cooperators cannot invade the residents.
In other words, the residents can distinguish the random cooperators from those following the social norm.

To conclude, the solitary observation model cannot sustain evolutionarily stable cooperation.
This is because of the independence of the opinions.
The conclusion remains valid for arbitrarily complex models, including non-binary reputation models, higher-order norms, and `dual-reputation update' models~\cite{murase2023indirect}.

The above argument is based on the assumption that there are at most two types of norms in the population, i.e., the mutation rate is sufficiently small.
We do not rule out the possibility that the cooperation can be stable with a mixture of multiple norms.
In previous studies for the solitary observation models, it was reported that coexistence between L3 and ALLC is stable against ALLD using the replicator equation~\cite{okada2018solution,okada2020two}.
Studies on the coexistence of multiple norms are still limited, and it is open whether the coexistence is stable against a wider range of norms.

\section{Analysis on the models with correlated opinions}

\subsection{Model formulation for the general case}

In the following, we discuss the optimal action for Alice when she has a similar opinion about Bob with Charlie.
Charlie considers Bob as $G$ with probability $h$. We consider the conditional probability $h_G$ that Alice also considers Bob as $G$ given Charlie considers Bob as $G$ as well.
Similarly, the conditional probability $h_B$ refers to the probability that Alice considers Bob as $G$ given Charlie considers Bob as $B$.

If Alice's opinion is statistically independent of Charlie's opinion, $h_G = h_B = h$ is satisfied.
If Alice's opinion is positively correlated with Charlie's, $h_G > h > h_B$ since Alice is more likely to consider Bob as $G$ when Charlie considers him as $G$ as well.
In the perfectly correlated case, namely for the public assessment model, $h_G = 1$ and $h_B = 0$.
The anti-correlated case does not happen since these three players are chosen randomly in each round.

Here, we do not introduce any further assumptions on the mechanisms of how their opinions are correlated.
Alice and Charlie could have observed Bob's behavior simultaneously, or they could have exchanged their opinions about Bob by gossiping, or a public institution could have disseminated the information about Bob to all players.
We just assume that they are statistically correlated while the expected value of the reputation $h$ is not biased by gossiping or whatever.

The probabilities $h$, $h_G$, and $h_B$ are not independent of each other.
Consider the probability that a pair of randomly chosen players have opposite opinions about a player Bob.
These players are written as $h(1-h_G)$ as well as $(1-h)h_B$. Since these must coincide, the following equality must hold:
\begin{equation}
  h(1-h_G) = (1-h) h_B
\label{eq:h_hG_hB}
\end{equation}
Obviously, when $h_G = h$, $h_B = h$ as well. Otherwise, $h_G > h > h_B$.

Next, consider the monomorphic population in which the players use the same norm $\{P, R\}$. The average reputation follows the following dynamics:
\begin{equation}
  \dot{h} = h h_G R_P(G, G) + h(1-h_G)     R_P(G, B) + (1-h)h_B     R_P(B, G) + (1-h)(1-h_B) R_P(B, B) -h
  \label{eq:h_dot_correlated}
\end{equation}
Using Eq.~(\ref{eq:h_hG_hB}), the following equation is satisfied in the stationary state:
\begin{equation}
  h = h h_G            R_P\lp{G, G}
    + h\lp{1-h_G}      \lb{ R_P\lp{G, B} + R_P\lp{B, G} }
    + \lp{1-2h + hh_G} R_P\lp{B, B}
\label{eq:h_ast_correlated}
\end{equation}
Therefore, given the norm $\{P, R\}$, $h$ is determined once $h_G$ is fixed, and vice versa.
There is only one degree of freedom.
The value of $h$ (or $h_G$) depends on the details of the model definitions.
Once we define the model to update the reputation (such as solitary observation, simultaneous observation, gossiping, or public assessment) and the norm that the population is using, $h$ as well as $h_G$ and $h_B$ are uniquely determined.
For the solitary observation model, where $h = h_G = h_B$, Eq.~(\ref{eq:h_ast_correlated}) is identical to Eq.~(\ref{eq:h_dot_solitary}).

Depending on the models, the values of $h$ and $h_G$ are not always obtained analytically, so we may need simulations to estimate these values.
(For instance, a rigorous analytic solution is yet to be found for the simultaneous observation model with observation probability $q < 1$.)
We can calculate $h$ and $h_G$ from simulations by measuring the average and variance of goodness~\cite{fujimoto2022reputation}.
Here, we define the goodness of player-$i$, $g_i$, as the fraction of the good image in the $i$-th column of the image matrix excluding the diagonal element:
\begin{equation}
g_i = \sum_{j \neq i} \delta(m_{ji}, G) / (N-1),
\end{equation}
where $\delta(x, y)$ is the Kronecker delta function.
The average goodness taken over $i$ equals $h$: $h = \langle g_i \rangle$.
The product $h h_G$ is the expected probability that two randomly chosen players agree that another randomly chosen player is $G$.
Thus, for a finite $N$,
\begin{equation}
\begin{split}
h h_G &= \left\langle g_i \frac{\lp{N-1}g_i - 1}{N-2} \right\rangle = \frac{N-1}{N-2} \left\langle g_i^2 \right\rangle - \frac{\left\langle g_i \right\rangle}{N-2}  \\
  h_G &= \frac{1}{N-2} \lb{ \lp{N-1} \frac{\left\langle g_i^2 \right\rangle}{\left\langle g_i \right\rangle} - 1 }.
\end{split}
\label{eq:h_hG_from_g}
\end{equation}
When $N \to \infty$, $h_G \to \langle g_i^2 \rangle / \langle g_i \rangle$.

\subsection{Invasion analysis}

Next, we consider a mutant having a different action rule $P'$ but the same assessment rule $R$.
We denote Alice's action rule as $P'(X)$ where $X$ is Alice's opinion about Bob.
We are going to show that $P(G) = 1$ and $P(B) = 0$ is Alice's best response to maximize her payoff under a certain condition for the resident social norm, $\{P, R\}$, $h$, and $h_G$.
Namely, Alice's best action rule is the deterministic discriminator under these cases. Otherwise, the best action rule is unconditional cooperator or defector.

We denote the expected opinion from a resident to the mutant as $H$. The dynamics of $H$ is written as
\begin{equation}
  \dot{H} = h h_G R_{P'}\lp{G,D} + h\lp{1-h_G} \lb{ R_{P'}\lp{G,D} + R_{P'}\lp{B,D} } + \lp{1-h} \lp{1-h_B} R_{P'}\lp{B,D} -H,
\end{equation}
where $R_{P'}\lp{X, Y} = P'\lp{X} R\lp{Y,C} + \lp{ 1-P'\lp{X} }R\lp{Y,D}$.
After a sufficiently long time, $H$ converges to its stationary value
\begin{equation}
  H^{\ast} = h h_G R_{P'}\lp{G,D} + h \lp{1-h_G} \lb{ R_{P'}\lp{G,D} + R_{P'}\lp{B,D} } + \lp{1-2h+hh_G} R_{P'}\lp{B,D},
\end{equation}
where Eq.~(\ref{eq:h_hG_hB}) is used to simplify the expression.
In the following, $H^{\ast}$ is denoted as $H$ as we only focus on the stationary state.
The cooperation probability from the residents to the mutant is
\begin{equation}
\begin{split}
  p_{\rm res \to mut} &= H P(G) + \lp{ 1-H } P(B)   \\
                      &=\quad  h h_G              \lc{ P'\lp{G} R\lp{G,C} + \lb{1-P'\lp{G}} R\lp{G,D} } P\lp{G}   \\
                      &\quad + h h_G              \lc{ P'\lp{G} \lb{1-R\lp{G,C}} + \lb{1-P'\lp{G}} \lb{1-R\lp{G,D}} } P\lp{B}    \\
                      &\quad + h\lp{1-h_G}        \lc{ P'\lp{B} R\lp{G,C} + \lb{1-P'\lp{B}} R\lp{G,D} } P\lp{G}  \\
                      &\quad + h\lp{1-h_G}        \lc{ P'\lp{B} \lb{1-R\lp{G,C}} + \lb{1-P'\lp{B}} \lb{1-R\lp{G,D}} } P\lp{B}   \\
                      &\quad + h\lp{1-h_G}        \lc{ P'\lp{G} R\lp{B,C} + \lb{1-P'\lp{G}} R\lp{B,D} } P\lp{G}   \\
                      &\quad + h\lp{1-h_G}        \lc{ P'\lp{G} \lb{1-R\lp{B,C}} + \lb{1-P'\lp{G}} \lb{1-R\lp{B,D}} } P\lp{B}  \\
                      &\quad + \lp{1-2h+hh_G}     \lc{ P'\lp{B} R\lp{B,C} + \lb{1-P'\lp{B}} R\lp{B,D} } P\lp{G}  \\
                      &\quad + \lp{1-2h+hh_G} \lc{ P'\lp{B} \lb{1-R\lp{B,C}} + \lb{1-P'\lp{B}} \lb{1-R\lp{B,D}} } P\lp{B}  \\
                      &\\
                      &=\quad h h_G P'\lp{G}          \lc{ R\lp{G,C} P\lp{G} - R\lp{G,D} P\lp{G} - R\lp{G,C} P\lp{B} + R\lp{G,D} P\lp{B} }  \\
                      &\quad+ h\lp{1-h_G} P'\lp{B}    \lc{ R\lp{G,C} P\lp{G} - R\lp{G,D} P\lp{G} - R\lp{G,C} P\lp{B} + R\lp{G,D} P\lp{B} }  \\
                      &\quad+ h\lp{1-h_G} P'\lp{G}    \lc{ R\lp{B,C} P\lp{G} - R\lp{B,D} P\lp{G} - R\lp{B,C} P\lp{B} + R\lp{B,D} P\lp{B} }  \\
                      &\quad+ \lp{1-2h+hh_G} P'\lp{B} \lc{ R\lp{B,C} P\lp{G} - R\lp{B,D} P\lp{G} - R\lp{B,C} P\lp{B} + R\lp{B,D} P\lp{B} }  \\
                      &\quad+ h h_G                   \lc{ R\lp{G,D} P\lp{G} + \lb{1-R\lp{G,D}} P\lp{B} }  \\
                      &\quad+ h\lp{1-h_G}             \lc{ R\lp{G,D} P\lp{G} + \lb{1-R\lp{G,D}} P\lp{B} }  \\
                      &\quad+ h\lp{1-h_G}             \lc{ R\lp{B,D} P\lp{G} + \lb{1-R\lp{B,D}} P\lp{B} }  \\
                      &\quad+ \lp{1-2h+hh_G}          \lc{ R\lp{B,D} P\lp{G} + \lb{1-R\lp{B,D}} P\lp{B} }  \\
                      &\\
                      &=\quad h h_G          P'\lp{G}  \lc{ \lb{R\lp{G,C} - R\lp{G,D}} \lb{P\lp{G} - P\lp{B}} }  \\
                      &\quad+ h\lp{1-h_G}    P'\lp{B}  \lc{ \lb{R\lp{G,C} - R\lp{G,D}} \lb{P\lp{G} - P\lp{B}} }  \\
                      &\quad+ h\lp{1-h_G}    P'\lp{G}  \lc{ \lb{R\lp{B,C} - R\lp{B,D}} \lb{P\lp{G} - P\lp{B}} }  \\
                      &\quad+ \lp{1-2h+hh_G} P'\lp{B}  \lc{ \lb{R\lp{B,C} - R\lp{B,D}} \lb{P\lp{G} - P\lp{B}} }  \\
                      &\quad+ h        \lc{ R\lp{G,D} P\lp{G} + \lb{1-R\lp{G,D}} P\lp{B} }  \\
                      &\quad+ \lp{1-h} \lc{ R\lp{B,D} P\lp{G} + \lb{1-R\lp{B,D}} P\lp{B} }  \\
                      &\\
                      % &= h     [R(G,C) - R(G,D)] [P(G) - P(B)] [ h_G P'(G) + (1-h_G) P'(B) ]   \\
                      % &+ (1-h) [R(B,C) - R(B,D)] [P(G) - P(B)] [ h_B P'(G) + (1-h_B) P'(B) ]   \\
                      % &+ h     [R(G,D) - R(B,D)] [P(G) - P(B)] + R(B,D)P(G) + [1-R(B,D)]P(B)   \\
                      % &\\
                      % &= h \Delta_G [ h_G P'(G) + (1-h_G) P'(B) ]   \\
                      % &+ (1-h) \Delta_B [ h_B P'(G) + (1-h_B) P'(B) ]  \\
                      % &+ h     [R(G,D) - R(B,D)] [P(G) - P(B)] + R(B,D)P(G) + [1-R(B,D)]P(B)
                      &=\quad  \lb{ hh_G \Delta_G + h\lp{ 1-h_G } \Delta_B } P'(G)   \\
                      &\quad + \lb{ h\lp{1-h_G} \Delta_G + \lp{1-2h+hh_G} \Delta_B } P'(B)  \\
                      &\quad  + h        \lc{ R\lp{G,D} P\lp{G} + \lb{1-R\lp{G,D}} P\lp{B} }
                              + \lp{1-h} \lc{ R\lp{B,D} P\lp{G} + \lb{1-R\lp{B,D}} P\lp{B} }
                      %&\quad + h     \lb{ R\lp{G,D} - R\lp{B,D} } \lb{ P\lp{G} - P\lp{B} } + R\lp{B,D}P\lp{G} + \lb{ 1-R\lp{B,D} } P\lp{B}
\end{split}
\end{equation}
where we defined
\begin{equation}
\begin{split}
  \Delta_G &\equiv \lb{ R\lp{G, C} - R\lp{G, D} } \lb{ P\lp{G} - P\lp{B} }  \\
  \Delta_B &\equiv \lb{ R\lp{B, C} - R\lp{B, D} } \lb{ P\lp{G} - P\lp{B} }.
\end{split}
\end{equation}
These are interpreted as the incentives for cooperation when the recipient is assessed as $G$ or $B$ by a resident observer, respectively, and are dependent on the resident's norm only.
The cooperation probability from the mutant to the residents is
\begin{equation}
  p_{\rm mut \to res} = h P'(G) + (1-h) P'(B)
\end{equation}
Therefore, the mutant's payoff is
\begin{equation}
\begin{split}
  \pi_{\rm mut} &= b p_{\rm res \to mut} - c p_{\rm mut \to res}    \\
                % &= h h_G ( b\Delta_G - c ) P'(G) + h(1-h_G) ( b\Delta_G - c ) P'(B) + (1-h) h_B ( b\Delta_B - c ) P'(G) + (1-h)(1-h_B) ( b\Delta_B - c ) P'(B)  \\
                % &+ b \{ h     [R(G,D) - R(B,D)] [P(G) - P(B)] + R(B,D)P(G) + [1-R(B,D)]P(B) \}  \\
                % &= [ h h_G ( b\Delta_G - c ) + (1-h)h_B ( b\Delta_B - c ) ] P'(G)   \\
                % &+ [ h(1-h_G) ( b\Delta_G - c ) + (1-h)(1-h_B) ( b\Delta_B - c ) ] P'(B)  \\
                % &+ \text{(terms that are independent of $P'$, $h_G$, and $h_B$)}  \\
                % &\\
                &= \alpha_G P'(G) + \alpha_B P'(B) + \text{(terms that are independent of $P'$, $h_G$, and $h_B$)},
\end{split}
\end{equation}
where we defined
\begin{equation}
\begin{split}
  \alpha_G &\equiv h h_G ( b\Delta_G - c ) + h(1-h_G) ( b\Delta_B - c )  \\
  \alpha_B &\equiv h(1-h_G) ( b\Delta_G - c ) + (1-2h+hh_G) ( b\Delta_B - c ).
\end{split}
\label{eq:alpha_G_B_definition}
\end{equation}
Since $\pi_{\rm mut}$ is a linear function of $P'(G)$ and $P'(B)$, the best action rule $\hat{P'}$ for the mutant is summarized as follows:
\begin{equation}
\begin{split}
  \hat{P'}(G) = \begin{cases}
    1  &  \text{when $\alpha_G > 0$}  \\
    \text{any} & \text{when $\alpha_G = 0$}  \\
    0  &  \text{when $\alpha_G < 0$}
    \end{cases}  \\
  \hat{P'}(B) = \begin{cases}
    1  &  \text{when $\alpha_B > 0$}  \\
    \text{any} & \text{when $\alpha_B = 0$}  \\
    0  &  \text{when $\alpha_B < 0$}
    \end{cases}.  \\
\end{split}
\label{eq:mut_best_action}
\end{equation}
From this equation, we conclude the best action rule $\hat{P'}$ is a deterministic action rule unless the special cases of $\alpha_G = 0$ or $\alpha_B = 0$.
When $\alpha_G$ and $\alpha_B$ are the same sign, ALLC or ALLD is the best action rule.
Conditional cooperation is the best action rule when $\alpha_G$ and $\alpha_B$ have the opposite signs.
(Because of the symmetry of $G$ and $B$, we only consider the case when $\alpha_G > 0$ and $\alpha_B < 0$ without loss of generality in the following.)

When the mutant cannot distinguish Charlie's opinion, namely, $h_G = h_B = h$, these two must have the same sign since $\alpha_G$ is a positive constant multiple of $\alpha_B$ as follows:
\begin{equation}
  \frac{\alpha_G}{h} = \frac{\alpha_B}{1-h} = h ( b\Delta_G - c ) + (1-h) ( b\Delta_B - c )
\end{equation}
Therefore, $\hat{P'}$ is either unconditional cooperation or defection in this case, confirming the conclusion in the previous section.

When $\alpha_G$ and $\alpha_B$ have the opposite signs, the deterministic conditional cooperation $P'(G) = 1$ and $P'(B) = 0$ is the best action.
To have $\alpha_G > 0$ and $\alpha_B < 0$, it is necessary that
\begin{equation}
  b\Delta_G - c > 0 \quad\text{and} \quad b\Delta_B - c < 0.
\end{equation}
The left-hand sides of these inequalities ($b\Delta_G - c$ and $b\Delta_B - c$) are the expected extra payoffs from cooperation when the observer's opinion about the recipient is $G$ and $B$, respectively.
If the residents are more able to detect defectors, $b\Delta_G - c$ is larger. If the residents justify the punishment more against bad recipients, $b\Delta_B - c$ gets lower (stronger incentives to defect against bad players.)

When these conditions are satisfied, in order to maximize $\pi_{\rm mut}$, $\alpha_G$ needs to be maximized as well.
Thus, mutants are able to obtain a larger payoff with larger $h_G$ and smaller $h_B$.
In other words, the mutant Alice should synchronize her opinions with Charlie's as much as possible.

\subsection{Critical correlation levels for Simple Standing and Stern Judging}

Here, we consider specific norms, such as Simple Standing (L3) and Stern Judging (L6), to get more insights.

For L3 norm, substituting $R_P(G, G) = R_P(G, B) = R_P(B, B) = 1 - \mu_a$ and $R_P(B, G) = \mu_a$ into Eq.~(\ref{eq:h_ast_correlated}) yields
\begin{equation}
  \lp{1-2\mu_a} h h_G - 2 \lp{1-\mu_a} h + \lp{1-\mu_a} = 0.
  \label{eq:L3_h_hG_eq}
\end{equation}
Thus, $h$ is determined as a function of $h_G$ as
\begin{equation}
  h = \frac{ 1-\mu_a }{ 2\lp{1-\mu_a} - \lp{1-2\mu_a}h_G }.
\end{equation}
This is an increasing function of $h_G$.
For the solitary observation model, where $h_G = h$, this equation is reduced to a quadratic equation for $h$.
The minimal value of $h_G$ is obtained as its solution:
\begin{equation}
  h = \frac{ 1 - \mu_a - \sqrt{\mu_a\lp{1-\mu_a}} }{1 - 2\mu_a}.
  \label{eq:L3_hG_min}
\end{equation}
For instance, $h = h_G = 0.875$ when $\mu_a = 0.02$ for the solitary observation model.

One of the stability condition $\alpha_G > 0$ is then
\begin{equation}
  \begin{split}
    hh_G \lp{ b\Delta_G - c } + h \lp{1-h_G} \lp{ b\Delta_B - c } > 0  \\
    h_G \lp{ 1 - 2\mu_a } b > c \\
    \boverc > \frac{1}{ \lp{1 - 2\mu_a} h_G }
  \end{split}
\end{equation}
using $\Delta_G = 1-2\mu_a$ and $\Delta_B = 0$.
The other stability condition $\alpha_B < 0$ yields
\begin{equation}
  \begin{split}
    h\lp{1-h_G} \lp{ b\Delta_G - c } + \lp{1-2h+hh_G} \lp{ b\Delta_B - c } < 0  \\
    \lp{1-2\mu_a}h(1-h_G)b < \lp{1-h}c  \\
    \boverc < \frac{ 1-\mu_a - \lp{1-2\mu_a}h_G }{ \lp{1-2\mu_a}\lp{1-\mu_a}\lp{1-h_G} }.
  \end{split}
\end{equation}

From Eq.~(\ref{eq:L3_h_hG_eq}), we can learn some properties of the norms in the small error limit even without further details of the model.
When $\mu_a \to 0$, Eq.~(\ref{eq:L3_h_hG_eq}) is reduced to
\begin{equation}
  h h_G - 2h + 1 = 0.
\end{equation}
By definition $h_G \geq h$, thus $h h_G - 2h + 1 = 0 \geq (h - 1)^2$.
Therefore, $h \to 1$ and $h_G \to 1$ for L3 in the errorless limit whether the model assumes solitary observation, simultaneous observation, or gossiping.
This shows that L3 is able to maintain a good reputation even for the private assessment model.

For L6 norm, substituting $R_P(G, G) = R_P(B, B) = 1 - \mu_a$ and $R_P(G, B) = R_P(B, G) = \mu_a$ into Eq.~(\ref{eq:h_ast_correlated}) yields
\begin{equation}
  2\lp{1-2\mu_a}hh_G + \lp{-3 + 4\mu_a}h + \lp{1-\mu_a} = 0.
  \label{eq:L6_h_hG_eq}
\end{equation}
Thus, the relationship between $h$ and $h_G$ is
\begin{equation}
  h = \frac{ 1-\mu_a }{ 1 + 2\lp{1-2\mu_a}\lp{1-h_G}}.
  \label{eq:L6_h_hG}
\end{equation}
From this equation, $h$ is an increasing function of $h_G$.
For the solitary observation model, \eqref{eq:L6_h_hG_eq} is reduced to a quadratic equation for $h$:
\begin{equation}
  \lp{2h-1} \lb{ \lp{1-2\mu_a}h - \lp{1-\mu_a} } = 0,
\end{equation}
whose solution is $h = 1/2$.

The stability condition $\alpha_G > 0$ is written using $\Delta_G = \Delta_B = 1-2\mu_a$ as
\begin{equation}
  \begin{split}
    hh_G \lp{ b\Delta_G - c } + h \lp{1-h_G} \lp{ b\Delta_B - c } > 0  \\
    2\lp{ 1 - 2\mu_a } \lp{2h_G -1} b > c \\
    \boverc > \frac{1}{ \lp{2h_G - 1}\lp{1 - 2\mu_a} }.
  \end{split}
  \label{eq:L6_alpha_G_positive}
\end{equation}
The other condition $\alpha_B < 0$ is
\begin{equation}
  \begin{split}
    h\lp{1-h_G} \lp{b\Delta_G - c} + \lp{1-2h+hh_G} \lp{b\Delta_B - c} > 0  \\
    b \lp{1-2\mu_a} \lp{ -2hh_G + 3h - 1 } < (1-h)c  \\
  \end{split}
\end{equation}
From Eq.~(\ref{eq:L6_h_hG_eq}), $hh_G = \lb{ \lp{3-4\mu_a}h - \lp{1-\mu_a} } / 2\lp{1-2\mu_a}$.
Therefore, the above inequality is equivalent to
\begin{equation}
  b \mu_a \lp{ 1 - 2h } < \lp{ 1-\mu_a } c.
\end{equation}
Since $h \geq 1/2$, the above inequality is always satisfied. L6 is always stable against ALLC.

\section{Application to specific models}

\subsection{Public assessment model}

First, we consider the public assessment model where opinions are perfectly synchronized, namely $h_G = 1$ and $h_B = 0$.
More specifically, we consider a model in which a randomly chosen observer assesses the donor.
The assessment is publicly announced to all players and all the players agree on the assessment about the donor.
This model is equivalent to the public assessment model studied in previous literature~\cite{ohtsuki2004should,ohtsuki2006leading,murase2023indirect} when every player has the identical assessment rule.

For the public assessment model, the necessary and sufficient condition for second-order norms to have a strict Nash equilibrium with full cooperation is theoretically obtained. According to Eq.~(44) in Ref.~\cite{murase2023indirect}, the condition is written as:
\begin{equation}
  \begin{cases}
    P(G) = 1 \\
    P(B) = 0 \\
    R(G,C) = 1 \\
    R(B,D) > 0 \\
    R(G,D) < 1 \\
    \boverc > \frac{1}{ 1 - R\lp{G,D} } \\
    R(B,C) \leq R(B,D) {\quad \text{or} \quad}
    \boverc < \frac{1}{ R\lp{B,C} - R\lp{B,D} }
  \end{cases}.
  \label{eq:second_order_cess}
\end{equation}
This is consistent with Eq.~(\ref{eq:mut_best_action}). Mutant's best action rule is the deterministic discriminator $P(G)=1$ and $P(B)=0$ when
\begin{equation}
\begin{cases}
  \alpha_G &= h (b \Delta_G - c) > 0   \\
  \alpha_B &= (1-h) (b \Delta_B - c) < 0
\end{cases}.
\end{equation}
Since $0 < h < 1$, $P$ is the unique best action rule when
\begin{equation}
\begin{cases}
  b \lb{ R\lp{G, C} - R\lp{G, D} } > c   \\
  b \lb{ R\lp{B, C} - R\lp{B, D} } < c
\end{cases},
\end{equation}
which agrees with the last three inequalities in Eq.~(\ref{eq:second_order_cess}).
Note that the other conditions in Eq.~(\ref{eq:second_order_cess}) came from the full-cooperation condition.

\subsection{Simultaneous observation model}

Next, we consider a model where multiple observers assess the donor simultaneously.
When the donor cooperates or defects against the recipient, the other players observe the donor's action with probability $q$.
In the limit of $q \to 1/N$, the model falls back to the solitary observation model.
This model has been studied theoretically and numerically~\cite{hilbe2018indirect,fujimoto2022reputation,fujimoto2023evolutionary,schmid2021evolution}.

An approximate analytical result is obtained for $q=1$~\cite{fujimoto2022reputation,fujimoto2023evolutionary,fujimoto2024leader}.
They analytically calculated the distribution of the goodness in the stationary state by approximating it as a superposition of Gaussian functions.
They concluded that Simple Standing (L3) norm promotes stable cooperation against other second-order norms for a range of $b/c$.
When the error rate $\mu_a \ll 1$ and the population size $N \to \infty$, the distribution of the goodness of the L3 residents is written as the sum of delta functions (See Fig.~3D in Ref.~\cite{fujimoto2023evolutionary}):
\begin{equation}
  p(g) = (1-2\mu_a) \delta\lp{ g - \lp{1 - \mu_a} } + \mu_a \delta\lp{ g - \lp{1 - 3 \mu_a} } + \mu_a \delta\lp{g - 2\mu_a} + O(\mu_a^2).
  \label{eq:goddness_dist_L3_simultaneous}
\end{equation}
From their results, L3 is stable against ALLC and ALLD for $1 < b/c < 2$ when the implementation error rate tends to zero~\cite{fujimoto2024leader}.

From Eq.~(\ref{eq:goddness_dist_L3_simultaneous}) and Eq.~(\ref{eq:h_hG_from_g}), we calculate $h$ and $h_G$ as
\begin{equation}
\begin{split}
  h &= \int_0^1 g p(g) dg = 1 - 2\mu_a + O(\mu_a^2), \\
  h_G &= \frac{ \int_0^1 g^2 p(g) dg }{ \int_0^1 g p(g) dg } = 1 - \mu_a + O(\mu_a^2).
  \label{eq:h_hG_simultaneous}
\end{split}
\end{equation}
These are consistent with Eq.~(\ref{eq:h_ast_correlated}), validating the calculation.
According to Eq.~(\ref{eq:mut_best_action}), the best action rule for the mutant is the deterministic discriminator when
\begin{equation}
\begin{cases}
  \alpha_G > 0 & \iff \frac{b}{c} > 1 + O(\mu_a)  \\
  \alpha_B < 0 & \iff \frac{b}{c} < 2 + O(\mu_a)
\end{cases},
\end{equation}
reproducing the results by Fujimoto and Ohtsuki~\cite{fujimoto2023evolutionary}.

An analogous calculation can be done for other norms. For Stern Judging (L6) norm, the goodness distribution has a single peak at $g = 1/2$~\cite{fujimoto2023evolutionary}:
\begin{equation}
  p(g) = \delta(g - 1/2),
  \label{eq:goddness_dist_L6_simultaneous}
\end{equation}
which yields $h = h_G = 1/2$.
Thus, we obtain $\alpha_G < 0$ and $\alpha_B < 0$, irrespective of $b$ and $c$.
Therefore, unconditional defection is the best action rule for the mutant, indicating that L6 cannot stabilize cooperation in this model.

\begin{figure}[h]
  \centering
  \includegraphics[width=0.9\textwidth]{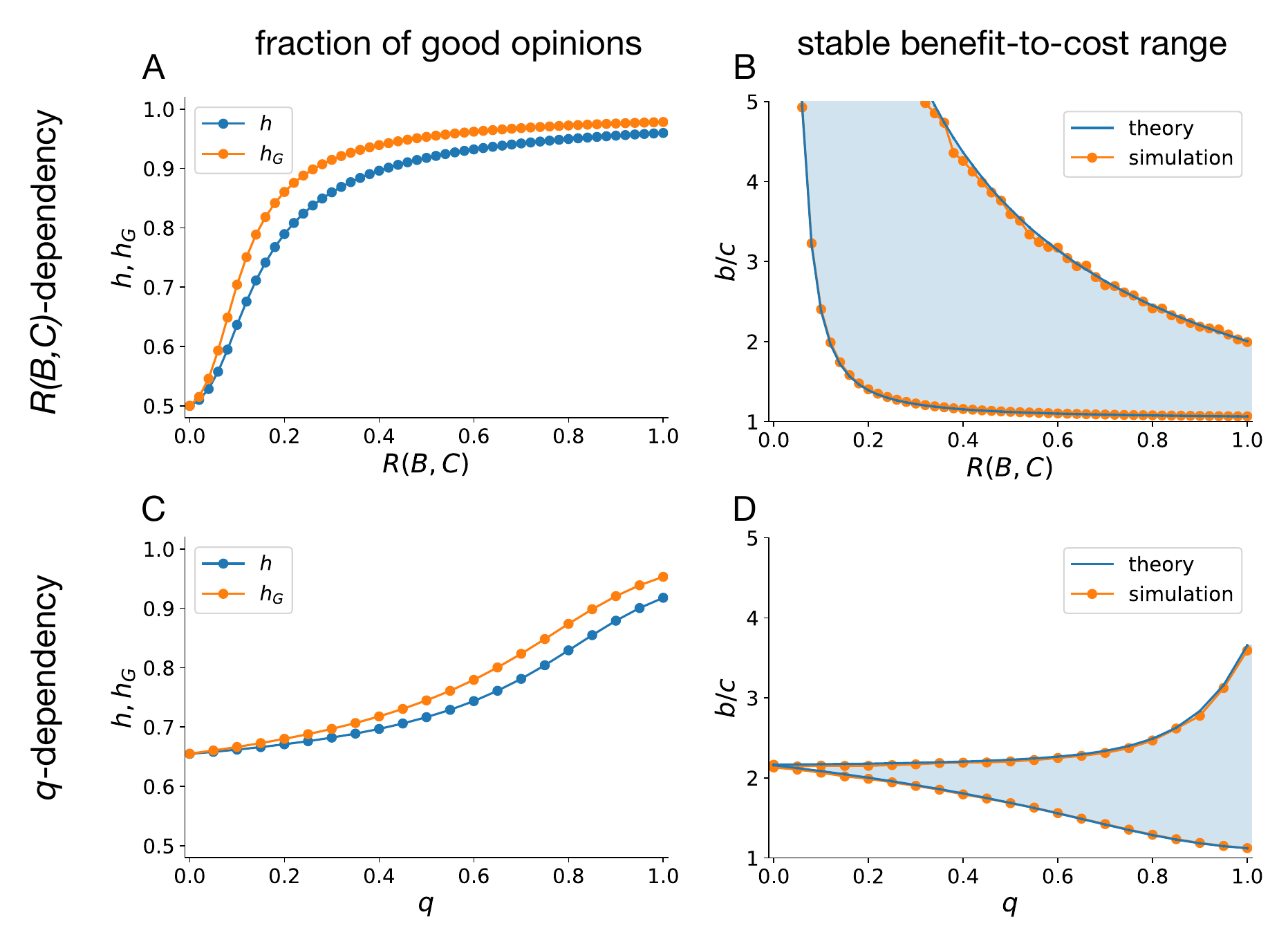}
  \caption{
    (A, B) An interpolation between the Simple Standing (L3) and Stern Judging (L6) norms for the simultaneous observation model.
    A norm is defined by $R(G, C) = R(B, D) = 1$, $R(G, D) = 0$, and $R(B, C)$ is controlled.
    (A) The relationships between $h$ and $h_G$ calculated by Monte Carlo simulation for monomorphic populations as functions of $R(B, C)$.
    (B) The right panel shows the stable range of $b/c$ for the discriminator strategy.
    The blue lines indicate the critical $b/c$ calculated by the theoretical conditions $\alpha_G > 0$ and $\alpha_B < 0$.
    To obtain these lines, we used $h$ and $h_G$ shown in (A).
    The red points indicate the same values calculated by Monte Carlo simulations by introducing a single ALLC/ALLD mutant.
    We introduced a single ALLC/ALLD mutant individual and calculated the average cooperation probabilities.
    We estimated the critical $b/c$ as the point where the payoff of the discriminator strategy is equal to that of the mutant.
    (C, D) The same plots as (A, B) but for different observation probability $q$.
    $R(B,C) = 0.5$ is fixed.
    As $q \to 0$, the results converge to the solitary observation model.
    The population size $N=100$ and the error rate $\mu_a = 0.02$. The results are averaged over $100$ independent runs.
  }
  \label{fig:interpolating_SS_SJ}
\end{figure}

In addition to these two norms, let us consider another set of norms that interpolates between L3 and L6, namely, $R(G, C) = R(B, D) = 1$, $R(G, D) = 0$, while $R(B, C)$ is varied in $[0, 1]$.
When $R(B, C) = 0$ and $R(B, C) = 1$, the norm is equivalent to L6 and L3, respectively.
Figure~\ref{fig:interpolating_SS_SJ}A shows $h$ and $h_G$ as functions of $R\lp{B, C}$.
Since no analytical solution exists for $h$ and $h_G$, we calculated these values using Monte Carlo simulations.
As $R(B, C)$ increases, both $h$ and $h_G$ increase quickly.
Even with $R(B, C) \approx 0.5$, $h$ goes as high as $0.9$, showing a high self-cooperation level.
Figure~\ref{fig:interpolating_SS_SJ}B shows the ranges of $b/c$ for which the discriminator action rule is stable.
Although no stable cooperation exists for L6, the lower bound of $b/c$ quickly drops as $R(B, C)$ increases, yielding a wide range of $b/c$ for stable cooperation.
As $R(B, C)$ approaches $1$, the upper bound of $b/c$ also decreases, narrowing the stable range.
L6 is too strict to stabilize cooperation, while L3 is too permissive to prevent invasion by ALLC.
In the intermediate region, the norm is more tolerant against ALLC while maintaining a reasonably high self-cooperation level.
These theoretical results are reproduced by Monte Carlo simulations, confirming the validity of our framework.

As another demonstration, we consider the effect of the observation probability $q$.
Figure~\ref{fig:interpolating_SS_SJ}C shows $h$ and $h_G$ as functions of $q$ for strategies with $R(G, C) = R(B, D) = 1$, $R(G, D) = 0$, and $R(B, C) = 0.5$.
As $q$ decreases to zero, $h$ and $h_G$ converge to the same value, reproducing the solitary observation model.
The stable range of $b/c$ for the discriminator strategy is shown in Fig.~\ref{fig:interpolating_SS_SJ}D.
Again, the stable range narrows as $q$ decreases and eventually vanishes when $q \to 0$, indicating that the solitary observation model cannot stabilize cooperation.

\subsection{Gossiping Model}

Another application of the above theory is the gossiping model proposed by Kawakatsu et al.~\cite{kawakatsu2024mechanistic}.
In their model, the gossiping process is introduced in the solitary observation model.
After a round of the donation games, observers assess the donors' actions as in the solitary observation model.
The gossiping process follows the private assessment.
Each player consults a randomly selected peer at a certain interval and adopts her view about the donor.
Donation games, private observations, and gossiping are repeated alternately until the reputations equilibrate.
The gossip dynamics for a focal individual are therefore described by a bi-allelic Wright-Fisher process, which keeps track of how many individuals view the focal individual as good or bad over discrete generations of gossip.

The number of gossip rounds quantifies the amount of peer-to-peer gossip between private observation periods.
The gossip duration $\tau$ quantifies the level of agreement between the players.
When $\tau \to 0$, the gossiping process is equivalent to the solitary observation model, while $\tau \to \infty$ corresponds to the public assessment model.

Here, we show a more detailed derivation of the critical gossip duration for the gossip model following the previous paper~\cite{kawakatsu2024mechanistic} with our notations.
As shown below, we obtain the two equalities of $h$ and $h_G$ by considering the reputation dynamics driven by assessments and gossiping.
These two equalities determine $h$ and $h_G$ after the equilibration of the reputations.
Then, we consider an infinitesimal amount of ALLD mutants and the stability of the discriminator strategy against the mutants by comparing their payoffs.
Hereafter, we only consider L6 (Stern Judging) norm with assessment error rate $\mu_a$, but analogous calculations can be done for other norms as shown in~\cite{kawakatsu2024mechanistic}.

% First, we consider the dynamics of $h$ driven by assessments.
% Plugging $R_P(G,C) = R_P(B,D) = 1-\mu_a$ and $R_P(G,D) = R_P(B,C) = \mu_a$ into Eq.~(\ref{eq:h_ast_correlated}), the relationship between the equilibrium reputation $h$ and its correlation $h_G$ is obtained as follows:
% \begin{equation}
%   \begin{split}
%     h &= hh_G (1-\mu_a) + 2h(1-h_G) \mu_a + (1-2h+hh_G) (1-\mu_a) \\
%     h &= -2h (1-2\mu_a) (1-h_G) + (1-\mu_a)  \\
%     h &= \frac{ 1-\mu_a }{ 1 + 2\lp{1-2\mu_a} \lp{1-h_G} }.
%   \end{split}
%   \label{eq:h_hG_relation_L6}
% \end{equation}
% Thus, $h$ is an increasing function of $h_G$.

We consider the reputation dynamics driven by gossiping.
By gossiping, the agreement level of the reputation increases while the average reputation remains unchanged.
According to Ref.~\cite{kawakatsu2024mechanistic}, the (dis)agreement levels after gossiping are updated according to the Wright-Fisher process.
They denote $\widetilde{g_2}$ ($\widetilde{b_2}$) as the probability that two randomly selected players agree that a third individual is $G$ ($B$) after gossiping.
These variables indicate the extent of agreement of the reputations.
They also introduced $2\widetilde{d_2}$, which is the probability that two randomly selected players have opposite opinions about a third individual.
By definition, $\widetilde{g_2} + \widetilde{b_2} + 2\widetilde{d_2} = 1$.
In our notation, these probabilities are expressed using $h$ and $h_G$ as follows:
\begin{equation}
\begin{split}
  \widetilde{g_2} &= h h_G\\
  \widetilde{b_2} &= (1 - 2h + hh_G)\\
  \widetilde{d_2} &= h(1-h_G).
\end{split}
\label{eq:gossip_agreement}
\end{equation}
Since these variables follow the Wright-Fisher process starting from the initial states $\widetilde{g_2} = h^2$, $\widetilde{b_2} = (1-h)^2$, and $\widetilde{d_2} = h(1-h)$, the agreement levels are updated as
\begin{equation}
  \begin{split}
  \widetilde{g_2} &= h^2 + h(1-h) \lp{ 1 - e^{-\tau} },  \\
  \widetilde{b_2} &= (1-h)^2 + h(1-h) \lp{ 1 - e^{-\tau} },  \\
  \widetilde{d_2} &= h(1-h)  e^{-\tau},
  \end{split}
  \label{eq:agreement_level_dynamics}
\end{equation}
where $\tau$ is the scaled gossip duration.
From these, we obtain a simple relationship between $h$ and $h_G$ after gossiping as
\begin{equation}
  1 - h_G = (1 - h) e^{-\tau}.
  \label{eq:h_hG_relation_gossip}
\end{equation}

Combining Eq.~(\ref{eq:h_hG_relation_gossip}) and Eq.~(\ref{eq:L6_h_hG}), we obtain the relationship between $\tau$ and $h$ as
\begin{equation}
  \begin{split}
  2\lp{1-2\mu_a} h(1-h) e^{-\tau} = 1 - h - \mu_a  \\
  e^{\tau} = \frac{ 2\lp{1-2\mu_a}h\lp{1-h} }{ 1-h-\mu_a }  \\
  \tau = \log \lb{ \frac{ 2\lp{1-2\mu_a}h\lp{1-h} }{ 1-h-\mu_a } }.
  \end{split}
  \label{eq:tau_h_relation}
\end{equation}
From the above, it is straightforward to show that $\tau$ is an increasing function of $h$ when $\mu_a \approx 0$.

Lastly, we consider the stability of the discriminator strategy against an infinitesimal amount of ALLD mutants.
The average reputation of the ALLD mutant is $\mu_a h + \lp{1-\mu_a} \lp{1-h}$ since it gets $G$ with probability $\mu_a$ and $1-\mu_a$ when it meets $G$ and $B$ recipients, respectively.
The payoff of an ALLD player is thus $b \lb{ \mu_a h + \lp{1-\mu_a} \lp{1-h} }$.
On the other hand, the residents' payoff is $\lp{b-c}h$.
The residents are stable if and only if
\begin{equation}
  \begin{split}
    b \lb{ \mu_a h + \lp{1-\mu_a} \lp{1-h} } &< (b - c)h,  \\
    %\boverc \lp{1 - \mu_a} \lp{ 2h-1 } > h   \\
    \lb{ 2\lp{\boverc}\lp{1-\mu_a} -1 }h &> \boverc\lp{1-\mu_a}  \\
    % h &> h^{\ast},
  \end{split}
  \label{eq:stability_condition_h_L6}
\end{equation}.
From this condition, Kawakatsu et al.~\cite{kawakatsu2024mechanistic} derived the critical gossip duration $\tau^{\ast}$ as
\begin{equation}
  \tau^{\ast} = \log \lb{
    \lp{ 2 - \frac{ \boverc }{ \boverc - \frac{1}{2\lp{1-\mu_a}} } } \lp{ \frac{ \boverc }{ \boverc - \frac{1}{1-2\mu_a} } }
  }.
\end{equation}

The same conclusion on the critical gossip duration is also obtained from our framework.
% For L6, $\Delta_G = 1 - 2\mu_a$ and $\Delta_B = -1 + 2\mu_a$.
For L6, the condition $\alpha_G > 0$ is calculated as in Eq.~(\ref{eq:L6_alpha_G_positive}).
%\begin{equation}
%  \begin{split}
%  \alpha_G > 0   \\
%  h_g \lb{ b\lp{1-2\mu_a} -c } + \lp{ 1-h_G } \lb{ -b\lp{1-2\mu_a} - c} > 0  \\
%  \lp{ 2h_G - 1} \lp{ 1-2\mu_a } b > c
%  \end{split}
%  \label{eq:alpha_G_positive_L6}
%\end{equation}
% From Eq.~(\ref{eq:L6_h_hG}), $h_G$ is obtained as a function of $h$:
% \begin{equation}
%   h_G = \frac{ 3h - 4\mu_a h - 1 + \mu_a }{ 2h\lp{1-2\mu_a} }.
% \end{equation}
% Therefore,
% \begin{equation}
%   2h_G - 1 = \frac{ \lp{2h-1} \lp{1-\mu_a} }{ h\lp{1-2\mu_a} }.
%  %&= \frac{ 3h - 4\mu_a h - 1 + \mu_a }{ 2h\lp{1-2\mu_a} } -1  \\
% \end{equation}
Plugging Eq.~(\ref{eq:L6_h_hG}) into Eq.~(\ref{eq:L6_alpha_G_positive}), we obtain
\begin{equation}
  \begin{split}
  \lp{2h-1} \lp{1-\mu_a} b &> ch  \\
  \lb{2\boverc\lp{1-\mu_a} - 1} h &> \boverc \lp{1-\mu_a},
  \end{split}
\end{equation}
which is identical to Eq.~(\ref{eq:stability_condition_h_L6}).
Thus, our theory is equivalent to the previous study in terms of the stability of the discriminator strategy.

\end{document}